\documentclass[useAMS,usenatbib]{mn2e}
\input psfig.tex
\usepackage{amsmath}

        \newcommand{\Hbeta}{\mbox{${\rm H_{\beta}}$}}
        \newcommand{\MgFe}{\mbox{${\rm [MgFe]} $}}
        \newcommand{\MgFei}{\mbox{${\rm [MgFe]'} $}}
        \newcommand{\MFe}{\mbox{${\rm \langle Fe \rangle} $}}
        \newcommand{\mgii}{\mbox{${\rm Mg_{2}} $}}
        
        \newcommand{\mgb}{\mbox{${\rm Mg_{b}} $}}
        \newcommand{\nad}{\mbox{${\rm NaD} $}}
        \newcommand{\cii}{\mbox{${\rm C_{2}4668} $}}

        \newcommand{\dHbeta}{\mbox{${\rm \delta H_{\beta}}$}}
        \newcommand{\dmgii}{\mbox{${\rm \delta Mg_{2}}$}}
        
        \newcommand{\dFe}{\mbox{${\rm \delta \langle Fe \rangle}$}}
        \newcommand{\dMgFe}{\mbox{${\rm \delta [MgFe]}$}}

    \newcommand{\asfe}{\mbox{{\rm $[\alpha/Fe]$}}}
    \newcommand{\alfa}{\mbox{$\alpha$-elements}}
    \newcommand{\alfe}{\mbox{$\alpha$-enhanced}}
    \newcommand{\enh}{\mbox{$\alpha$-enhancement}}
    
    \newcommand{\Msun}{\mbox{${\rm M_{\odot}}$}}   
    \newcommand{\Teff}{\mbox{${\rm T\sub{eff}}$}}
    
    \newcommand{\Lsun}{\mbox{${\rm L_{\odot}}$}}   

        \newcommand{\sub}[1]{\mbox{$_{\rm #1}$}}

        \def\smallskip{\vskip 4pt}

\title[Star Formation History in Early-Type Galaxies]{Star Formation History in Early-Type Galaxies.\\
I. The Line Absorption Indices Diagnostics}

\author[R. Tantalo \& C. Chiosi]{Rosaria Tantalo \& Cesare Chiosi\\
 Department of Astronomy, University of Padova,
       Vicolo dell'Osservatorio 2, 35122 Padova, Italy\\
E-mail: {\tt tantalo@pd.astro.it; chiosi@pd.astro.it}
}

\date{\tt Submitted: November 2003; Revised: April 2004}

\pubyear{2004}

\begin{document}
\maketitle
\title{Star Formation History in Early-Type Galaxies.\\
       I. The Line Absorption Indices Diagnostics}

\begin{abstract}
To unravel the formation mechanism and the evolutionary history of
Elliptical Galaxies (EGs) is one of the goals of modern
astrophysics. In a simplified picture of the issue, the question
to be answered is whether they have formed by hierarchical merging
of pre-existing sub-structures (maybe disc galaxies) made of stars
and gas, each merging event likely accompanied by strong star
formation, or conversely, they originated from the early
aggregation of lumps of gas turned into stars in the remote past
via a burst-like episode ever since followed by quiescence so as
to mimic a sort of monolithic process. Even if the two
alternatives seem to oppose each other, actually they may concur
to shaping the final properties of EG's as seen today. Are there
distinct signatures of the underlying dominant process in the
observational data? To this aim we have examined the line
absorption indices on the Lick system of the normal, field EGs of
\citet{Trager97} and the interacting EGs (pair- and shell-objects)
of \citet{Longhetti20}. The data show that both normal, field and
interacting galaxies have the same scattered but smooth distribution
in the \Hbeta\ vs. \MgFe\ plane even if the interacting ones show a
more pronounced tail toward high \Hbeta\ values. This may suggest that
a common physical cause is at the origin of their distribution. There
are two straightforward interpretations of increasing complexity: (1)
EGs span true large ranges of ages and metallicities. The age youth is
the signature of the aggregation mechanism, each event accompanied by
metal enrichment. This simple scheme cannot, however, explain other
spectro-photometric properties of EGs and has to be discarded. (2) The
bulk population of stars is old but subsequent episodes of star
formation scatter the EGs in the diagnostic planes. However, this
scheme would predict an outstanding clump at low \Hbeta\ values,
contrary to what is observed. The model can be cured by supposing that
the primary star formation activity lasted for a significant fraction
of the Hubble time (5\,Gyr$\leq$T$\leq$13\,Gyr) accompanied by global
metal enrichment. The ``younger'' galaxies are more metal-rich.  The
later burst of star formation should be small otherwise too many high
\Hbeta\ objects would be observed. Therefore, the distribution of normal, pair- and shell-galaxies in the \Hbeta\
vs. \MgFe\ plane is due to the global metal enrichment. Even though
the above schemes provide a formal explanation, they seem to be too
demanding because of the many {\it ad hoc} ingredients that have to be
introduced. Furthermore they neglect the observationally grounded hint
that the stellar content of EGs is likely enhanced in \alfa\ with
\asfe\ ranging from 0.1 to 0.4\,dex. We propose here a new scheme, in
which the bulk dispersion of galaxies in the \Hbeta\ vs. \MgFe\ plane
is caused by a different mean degree of enhancement.  In this model,
neither large age ranges nor universal enrichment law for the old
component are required and the observed distribution along
\Hbeta\ is naturally recovered. Furthermore, later bursts of stellar
activity are a rare event interesting only those galaxies with very
high \Hbeta\ (roughly $>2.5$). Finally, simulations of the scatter in
broad-band colors of EG's seem to confirm that the bulk stars have
formed in the remote past, and that mergers and companion star
formation in a recent past are not likely, unless the intensity of
the secondary activity is very small.
\end{abstract}


\section{Introduction}\label{intro}

Determining the age of the bulk stellar content of Elliptical
Galaxies (EGs) is basic to reconstructing the past history of star
formation and to setting clues on the galaxy formation process
itself. Two competing scenarios are proposed: the monolithic
picture in which a single dominant episode of star formation
occurred in the past, ever since followed by quiescence and
passive evolution (secondary episodes are however always
possible); the hierarchical scheme, in which a series of mergers
of smaller sub-units has taken place over the billions of years,
each of those likely accompanied by star forming activity (mergers
of inert objects with no additional star formation cannot be
excluded). Current ideas concerning the two competing pictures of
galaxy formation have been critically reviewed by \citet{Ellis98}
and \citet{Peebles02}. In short the present day view of structure
formation suggests that roughly half of the large EGs were
assembled at $z<1$. However, observational evidence of variations
with redshift of the color-magnitude, fundamental plane,
size-magnitude, K-magnitude-redshift relations indicate that large
EGs have already formed in the distant past at $z\geq 2$.

In order to discriminate between the two schemes, the key question
to be addressed and hopefully answered is what kind of signatures
the stellar populations of a galaxy have inherited from the
forming mechanism. The diagnostic to our disposal is mainly based
on magnitudes, colors, integrated spectra, luminosity weighed line
strength indices, estimates of chemical abundances and abundance
ratios together with the spatial gradients in those quantities. As
for the line absorption indices, gradients in \mgii\ and \MFe\
(and others) observed in EGs
\citep{Worthey92,Gonzalez93,Davies93,Carollo93,Carollo94a,Carollo94b,
Balcells94,Fisher95,Carrasco95,Fisher96} have often been interpreted
as indicating that the abundances of \alfa\ (Mg, O, etc.) with respect
to iron are enhanced -- \asfe$>$0 -- in the central regions. Opposite
conclusions however exist \citep{Kuntschner98,Davies2001}. Furthermore, 
limited to the nuclear regions, indices vary passing from one galaxy
to another \citep{Gonzalez93,Trager20a,Trager20b}. Looking at the
correlation between \mgii\ and \MFe\ (or similar indices) for the
galaxies in those samples, \mgii\ increases faster that \MFe, which is
once more interpreted as due to enhancement of \alfa\ in some
galaxies. In addition to this, since the classical paper by
\citet{Burstein88}, the index \mgii\ is known to increase with the
velocity dispersion (and hence mass and luminosity) of the galaxy.
Standing on this body of data the conviction arose that the degree
of enhancement in \alfa\ ought to increase passing from dwarf to
massive EGs
\citep{Faber92,Worthey94,Matteucci94,Matteucci97,Matteucci98}.

Another point to consider is whether normal and interacting
galaxies (i.e. shell- and pair-objects) show any systematic
difference in colors and line absorption indices. The subject was
addressed by
\citet{Longhetti98a,Longhetti98b,Longhetti99,Longhetti20}, whose
results will be shortly summarized below.

Using this body of data, can we infer the age, the metallicity and
the degree of enhancement for the bulk stars in an EG? The task is
not trivial because age and metallicity have similar effects on
the spectrum of a galaxy, i.e. the spectrum and hence the colors
of an old and metal-poor population may look like those of a young
and metal-rich one, the so-called {\it age metallicity-degeneracy}
pointed out long ago by \citet{Renbuz86}. The degeneracy is also
further complicated by effects both on the age and metallicity
caused by the enhancement in $\alpha$-elements. A promising
way-out is perhaps offered by the system of line absorption
indices introduced by the Lick group \citep[][and reference
therein]{Worthey92,Worthey94} which have been extensively used to
infer the age, metallicity, and abundance ratios of EGs
\citep{Bressan96,Tantalo98a,Kuntschner98a,Kuntschner98,
Jorgensen99,Kuntschner00,Trager20a,Trager20b,Kuntschner01b,Vazdekis01,
Davies2001,Poggianti01,Maraston03,Thomas03,Thomas03a,ThoMara03,Tantalo04a}.

Let us shortly summarize the key steps of those articulated
analysis, paying particular attention to the age for which often
the conclusion that EGs show evidence of recent star formation
either for the whole bulk stellar population or part it has been
drawn. The age youth has often been taken as the signature of the
hierarchical mechanism.

(i) Starting from the pioneering study of \citet{Gonzalez93} of
``normal galaxies'' (i.e. those with no sign of dynamical
interaction, accretion, etc.) in the local Universe, different
groups analyzed the distribution of galaxies in the \Hbeta\ vs.
\MgFe\ diagnostic plane and others of the same kind
\citep{Buzzoni92,Buzzoni94,Bressan96,Tantalo98,Tantalo98a} using the Single
Stellar Populations (SSP) approximation, i.e. the stellar content
of a galaxy is reduced to SSP of suitable age, metallicity and
degree of enhancement. The index \Hbeta\ is considered to be a
good age indicator, whereas \MgFe\ is considered to be most
sensitive to the metallicity. In reality both indices are
sensitive to age, metallicity, and degree of enhancement
\citep[see also][ for similar remarks]{Gonzalez93,Tantalo04a}.
Since EGs are more scattered in \Hbeta\ than in \MgFe\ and do not
follow the relation expected for coeval old objects matching the
Color-Magnitude relation \citep{Bower92a,Bower92b}, it was argued
that \Hbeta\ traces the age of the bulk stars, thus suggesting
that some galaxies are truly young objects. However, a closer
scrutiny of the problem led \citet{Bressan96} to suggest all
galaxies in that sample should be old but with different histories
of star formation. Some of them completed their stellar activity
in the far past with no evidence of subsequent episodes. Others
had a more prolonged star forming history, perhaps in recurrent
episodes of short duration. Looking at the galaxy to galaxy
differences \dHbeta\ and \dMgFe\ in the nuclear values and their
correlation with $\Sigma_{0}$, the suggestion was advanced that
the global duration of the star forming activity gets longer at
decreasing $\Sigma_{0}$ (galaxy mass). This was later confirmed by
dynamical Tree-SPH simulations of early type galaxies by
\citet{Kawata99,Kawata01,Kawata01a} and \citet{Chiocar02}. In
addition to this, \citet{Bressan96} and \citet{Longhetti20} tried
to understand the effect brought to the indices of a population of
old stars by a recent episode of star formation. The results is
that indices like \Hbeta\ are strongly affected by even small
percentages of young stars: as long as star formation is active
they jump to very high values and when star formation is over they
fall back to the original value on a time-scale of about 1\,Gyr.
Other indices like \mgii, \MFe\ are much less affected even if the
companion chemical enrichment may somewhat change them. In
diagnostics planes like \Hbeta\ vs. \MFe\ the galaxy performs and
extended loop elongated towards the \Hbeta\ axis, thus causing an
artificial dispersion which could be interpreted as an age
dispersion, whereas what we are really seeing is the transient
phase associated to the temporary stellar activity. Consequently,
the idea that the dispersion in \Hbeta\ measures the age of the
last episode of star formation instead of the age of the bulk
population was commonly accepted. Finally, EGs likely have mean
metallicities in the range $Z_{\odot} \leq Z \leq 3Z_{\odot}$ as
indicated by comparing theoretical models with data \citep[see
also][]{Bressan96,Greggio96}.

(ii) \citet{Tantalo98a} examining the Gonz\'alez galaxies, in
order to infer from the simultaneous fit of \Hbeta, \mgii, and
\MFe\ the age, the metallicity, and the enhancement factor \asfe,
noticed that: (a) In a small group of galaxies, the nucleus
appears to be more metal-rich, more \alfe, and younger (i.e.
containing a significant fraction of young stars) than the
external regions. (b) The galaxy to galaxy differences \dHbeta,
\dmgii, and \dFe\ for the nuclear regions yield age differences
for the last episode of star formation perhaps correlating with
the galaxy mass. In other words, while there seems to be a sort of
{\it upper limit} to the age of the bulk stars, traced by
quiescent galaxies (no signs of rejuvenation), is seems also that
the last episode of star formation took place closer and closer to
the present at decreasing galaxy luminosity (mass).

(iii) Surprisingly, in the \Hbeta\ vs. \MgFe\ plane, shell-,
pair-galaxies (i.e. those with signs of dynamical interactions)
and normal galaxies share the same distribution
\citep{Longhetti20}. There is, however, a group of peculiar
galaxies with much stronger \Hbeta\ as compared to the normal
ones. Does it mean that the scatter seen in this diagram has a
common origin, perhaps secondary episodes of star formation that
can occur independently of whether or not a galaxy is interacting?
This will be discussed in detail in Section~\ref{random_bursts}
below.

Thanks to these achievements, the \Hbeta\ vs. \MgFe\ plane (and
others of the same kind) for SSP of both solar and
$\alpha$-enhanced abundance ratios soon became a popular tool to
estimate the age, metallicity and degree of enhancement of
galaxies \citep{Kuntschner98a,Kuntschner98,Jorgensen99,
Kuntschner00,Kuntschner01b,Vazdekis01,Davies2001,Poggianti01}. The
advent of large libraries of SSPs with different chemical
compositions including the effect of different \asfe\ ratios, e.g.
\citet{Salasnich20}, on one hand spurred the construction of
corresponding libraries of line absorption indices, on the other
hand made the many-indices fitting technique (three at least) of
general use \citep[e.g.][]{Trager20a,Trager20b,Maraston03,
Thomas03,Thomas03a,ThoMara03,Tantalo04a} and introduced new
techniques more sophisticated than the two-indices diagnostics.
The solution for ages, metallicities and \asfe\ ratios based on
large samples of galaxies, e.g. the \citet{Trager97} list, once
more yields large ranges of ages, metallicities, and abundance
ratios, as amply discussed by \citet{Tantalo04a}.

Although we have repeatedly mentioned that the large spread in age
formally assigned to galaxies (EGs in particular) on the base of
the line absorption diagnostics has to be interpreted as referring
to the last episode of star formation rather than the age of the
bulk stellar population, quantifying the relative efficiency of
the different star forming processes has still remained unsettled
but for a few preliminary studies \citep{Bressan96,Longhetti20}.
The question is: how much a recent, minute episode of star
formation, engaging a small fraction of the total mass, may alter
the indices of an otherwise old population of stars in EGs? In
this paper we reconsider the whole problem at the light of recent
developments in the observational data and theory of population
synthesis concerning the theoretical line absorption indices in
which the effect of \asfe\ ratios are taken into account. We
intend in particular to investigate the effect of different
histories of star formation and chemical enrichment from galaxy to
galaxy and to explore in more detail the effects of late bursts of
stellar activity.

The paper is organized as follows. In Section~\ref{indices} we
shortly summarize the definition of line absorption indices and
total enhancement parameter, mention the {\it Fitting Functions}
in use, recall the main properties of stellar models and
associated isochrones we have adopted to calculate SSPs and their
integrated indices, and finally describe the key ingredients of
the library of stellar spectra we have used to calculate indices
and broad-band colors of SSPs. In Section~\ref{ind_ssp} we
describe the indices for SSPs at varying metallicity, enhancement
parameter, and age. In Section~\ref{composing} we present the
technique to calculate the indices of composed objects made of at
least two SSPs of different metallicity, enhancement, and age. In
addition to this, we show the path in several two-indices-plane
described by old galaxies rejuvenated by an episode of star
formation of different intensity and age. In
Section~\ref{sim_ssp_age}, we present the observational data we
want to interpret, i.e. the \citet{Trager97} sample of normal
galaxies complemented by the sample of \citet{Longhetti20} for
shell- and pair-galaxies. The analysis is based on the
distribution of galaxies in the diagnostic plane \Hbeta\ vs.
\MgFe. Three different interpretations are explored: (i) The bulk
stellar population of early-type galaxies span large ranges of
metallicities and ages. This is equivalent to say that galaxies
can form over time interval comparable to the Hubble time. (ii)
The bulk stellar population is very old (say 10-13\,Gyr) in all
galaxies but recent episodes of star formations (bursts) engaging
a small fraction of the total mass are present in most of them.
Both explanations can hardly fit the distribution of
galaxies along the \Hbeta\ direction unless {\it ad hoc}
hypotheses are made for a sort of Universal Law of Metal
Enrichment (ULME). Since severe drawbacks of these explanations
are either the large range of ages together with the {\it ad hoc}
hypothesis of ULME and/or the otherwise unavoidable existence of
subsequent bursts of stellar activity in all galaxies, a third
alternative is explored. (iii) The bulk population of stars in all
galaxies is very old (once again 10-13\,Gyr) but span large ranges
of metallicity and global enhancement factor. This is equivalent
to say that each galaxy has its own history of star formation
determining the chemical properties. No ULME and no subsequent
bursts of stellar activity are required to account for the
observed distribution of the majority of galaxies. Only a few
objects with unusually strong \Hbeta\ are likely in the bursting
mode. In Section~\ref{discussion}, the last interpretation is
critically discussed, the case of two template galaxies is
examined in detail, and a few remarks are made on the different
range of \Hbeta\ spanned by EGs in Coma and local vicinity. In
Section~\ref{colors} we address the question: how intense and old
subsequent bursts of stellar activity ought to be without
violating the constraint imposed by the present-day mean
broad-band colors of EGs? Finally, in Section~\ref{concl} we
summarize the results and presents some concluding remarks.

\section{Theory of line absorption indices}\label{indices}

\subsection{Definition}\label{def_ind}

The technique to calculate line-strength indices of SSPs is amply
described in \citet{Worthey94,Bressan96,Tantalo98a,Maraston03} and
\citet{Tantalo04a}. The reader is referred to those articles and
references therein for a detailed description of the method. Here
we limit ourselves to summarize a few basic points for the sake of
clarity. Suffice here to recall that a line absorption index is
constructed from the ratio $F_{l}/F_{c}$ where $F_{l}$ and $F_{c}$
are the fluxes in the line and pseudo-continuum, respectively. The
flux $F_{c}$ is calculated by interpolating to the central
wavelength of the absorption line, the fluxes in the midpoints of
the red and blue pseudo-continua bracketing the line
\citep{Worthey94}.

\subsection{Fitting Functions}\label{fittff}

Since the Lick system of indices \citep{Burstein84,Faber85,
Worthey94} stands on a spectra library with fixed resolution of
8\AA, whereas most of the synthetic spectra in use have a
different resolution, to overcome the difficulty, the so-called
{\it Fitting Functions} ($\mathcal{FF}$) have been introduced.
They express the indices measured on the observed spectra of a
large number of stars with known gravity, \Teff, and chemical
composition ([Fe/H]) as functions of these parameters
\citep{Worthey94}. In the following we adopt the \citet{Worthey92}
$\mathcal{FF}$, extended however to high temperature stars
(\Teff$\approx$10,000\,K) as reported in \citet{Longhetti98a}.

\subsection{Integrated indices of SSPs}

They are derived in the following way. A SSP is described by an
isochrone in the HRD whose elemental bins (fixed by $\Delta
\log{{\rm L}/\Lsun}$ and $\Delta \log{\Teff}$) can be conceived as
a ``star'' of suitable luminosity (gravity), \Teff\ and
metallicity. Each elemental bin is in turn populated by a number
$N_{i}$ of such stars given by

\begin{equation}
N_{i}=\int_{m_{a}}^{m_{b}}\phi(m)dm
\end{equation}

\noindent where $m_{a}$ and $m_{b}$ are the minimum and maximum
star mass in the bin and $\phi(m)$ is the initial mass function in
number. We start from the typical star of a bin for which we may
derive the indices from the $\mathcal{FF}$. The indices are then
inverted to derive the flux in the absorption line, $F^{*}_{l,i}$,
and in the pseudo-continuum $F^{*}_{c,i}$. Known the index for a
single star, we weigh its contribution to the integrated value on
the relative number of stars of the same type. We calculate the
ratio

\begin{equation}
\frac{\sum_{i} F^{*}_{l,i} N_{i}}
             {\sum_{i} F^{*}_{c,i} N_{i}}
\label{int_ssp}
\end{equation}

\noindent where $N_{i}$ is the above number of stars in the
generic bin, and finally insert the ratio into the definition of
each index as appropriate.

\subsection{$\alpha$-enhanced chemical compositions}\label{enha}

{\bf Total enhancement}. To distinguish the enhancement of
individual species from the {\it total one } characterizing a
given chemical mixture, \citet{Tantalo04a} introduce the parameter
$\Gamma$ defined as follows. Let us take a certain mixture of
elements with mass abundances $X_{j}$ of each species, and total
metallicity $Z$ (sum of the $X_{j}$ for all elements heavier than
He). The ratio of the mass abundance of an element with respect to
Fe and to the Sun is given by

\begin{equation}
\left[\frac{X_{j}}{X_{Fe}} \right] = log \left( \frac{X_{j}}{X_{Fe}} \right)
- log \left( \frac{X_{j}}{X_{Fe}} \right)_{\odot}
\end{equation}

\noindent Let us now assume that the ratio $[X_{j}/X_{Fe}]$ of
some elements is changed under the conditions that the abundance
of Fe with respect to the Sun and the total metallicity $Z$ remain
constant. The mass-abundances of all species in the new mixture
are accordingly scaled to a new value $X'_{j}$. The {\it total
enhancement factor} $\Gamma$ is

\begin{equation}
\Gamma = - \log{ \left( \frac{X'_{Fe}}{X_{Fe}^{\odot}} \right) }
\end{equation}

\noindent
See \citet{Tantalo04a} for all other details.

\noindent {\bf The \citet{Tripicco95} Response Functions}. A
method designed to include the effects of enhancement on line
absorption indices has been suggested by \citet[
TB95]{Tripicco95}, who introduce the concept of {\it Response
Functions}. In brief from model atmospheres and spectra for three
stars of assigned effective temperature and gravity, i.e. a
Cool-Dwarf (CD), a Turn-Off (TO), and a Cool-Giant (CG), they
calculate the absolute indices $I_{0}$. Doubling the abundances
$X_{i}$ of the C, N, O, Mg, Fe, Ca, Na, Si, Cr, and Ti in steps of
$\Delta [X_{i}/H]$=0.3\,dex they determine the incremental ratios
$\Delta I_{0}/\Delta [X_{i}/H]$ from which the {\it Response
Function} $R_{0.3}(i)$ for any index corresponding to a variation
of the element {\it i-th} $\Delta [X_{i}/H]=+0.3$\,dex, can be
derived

\begin{displaymath}
R_{0.3}(i) = \frac{1}{I_{0}}\, \frac{\Delta I_{0}}{\Delta [X_{i}/H]}\, 0.3
\end{displaymath}

\noindent The {\it Response Functions} for the Cool-Dwarf,
Turn-Off, and Cool-Giant stars constitute the milestones of the
calibration.

The mathematical algorithm for correcting an index from solar to
$\alpha$-enhanced element partitions is taken from
\citet{Trager20a} and \citet{Tantalo04a} to whom the reader should
refer for all details. In brief, starting from the assumption that
fractional variation of an index to changes of the chemical
parameters is the same as that for the reference index $I_{0}$,
the following expression is derived

\begin{equation}
 {\frac{\Delta I}{I}} = {\frac{\Delta I_0}{I_0}} =
 \left\{ \prod_{i} [1 + R_{0.3}(i)]^\frac{ [X_{i}/H]}{0.3} \right\} -1
\label{dind}
\end{equation}

\noindent where $R_{0.3}(i)$ are the {\it Response Function}
tabulated by TB95. Relation~(\ref{dind}) can be applied under the
obvious condition that $I/I_{0} > 0$.

The above algorithm is used to evaluate the fractional variations for
the three calibrators, i.e. $(\Delta I/I)_{CD}$, $(\Delta I/I)_{TO}$
and $(\Delta I/I)_{CG}$. Since in general a star or elemental bin
along an isochrone will have effective temperature and gravity
different from the ones of the three calibrators, and the fractional
variations for these latter are also different, to evaluate the total
fractional variation to be used for the particular star (isochrone
bin) under examination, we linearly interpolate both in effective
temperature and gravity among the fractional variations of the
calibrators weighing their contribution according to their distance
from the current star in the HR-Diagram. For a detailed discussion of
this topic and associated uncertainties see \citet{Tantalo04a}.

\begin{table}
\small
\begin{center}
\caption[]{Chemical composition and [Fe/H] as function of $\Gamma$ for the SSPs in use.}
\label{enh-deg}
\begin{tabular*}{82.5mm}{|c c c | r| r| r|}
\hline
\multicolumn{1}{|c}{} &
\multicolumn{1}{c}{} &
\multicolumn{1}{c|}{} &
\multicolumn{1}{c|}{$\Gamma=0.$} &
\multicolumn{1}{c|}{$\Gamma=0.3557$} &
\multicolumn{1}{c|}{$\Gamma=0.50$}\\
\hline
\multicolumn{1}{|c}{$Z$} &
\multicolumn{1}{c}{$Y$} &
\multicolumn{1}{c|}{$X$} &
\multicolumn{1}{c|}{[Fe/H]} &
\multicolumn{1}{c|}{[Fe/H]} &
\multicolumn{1}{c|}{[Fe/H]} \\
\hline
 0.008& 0.248 & 0.7440 & --0.3972 & --0.7529 &--0.8972\\
 0.019& 0.273 & 0.7080 &   0.0000 & --0.3557 &--0.5000\\
 0.040& 0.320 & 0.6400 &   0.3672 &   0.0115 &--0.1328\\
 0.070& 0.338 & 0.5430 &   0.6824 &   0.3267 &  0.1715\\
\hline
\end{tabular*}
\end{center}
\end{table}

\subsection{Stellar models and isochrones}\label{mod_iso}

The SSP indices are based on the Padova Library of stellar models
and companion isochrones according to the version by
\citet{Girardi20} and Girardi (2003, private communication). This
particular set of stellar models/isochrones differs from the
classical one by \citet{Bertelli94} for the efficiency of
convective overshooting and the prescription for the mass-loss
rate along the Asymptotic Red Giant Branch (AGB) phase. For the
reasons amply explained by \citet{Tantalo04a}, we prefer not to
use the more recent stellar models by \citet{Salasnich20} in which
the effect of \enh\ is already included in the stellar opacity.
Since indices are essentially a surface phenomenon in the sense
that they are derived from the $\mathcal{FF}$ linked to the
stellar models only via the effective temperature, surface
gravity, and iron content [Fe/H], details of the stellar models
caused by patterns of abundances enhanced in \alfa\ are of minor
relevance. The point has been made clear by the systematic
analysis of the issue made by \citet{Tantalo04a}. Therefore, the
stellar models and companion SSPs by \citet{Girardi20} and Girardi
(2003, private communication) are fully adequate to the purposes
of the present study.

The stellar models extend from the ZAMS up to either the start of the
thermally pulsing AGB phase (TP-AGB) or carbon ignition. No details on
the stellar models are given here; they can be found in
\citet{Girardi20} and \citet{Girardi02}. Suffice it to mention that:
(i) in low mass stars passing from the tip of red giant branch (T-RGB)
to the HB or clump, mass-loss by stellar winds is included according
to the \citet{Reimers75} rate with $\eta$=0.45; (ii) the whole TP-AGB
phase is included in the isochrones with ages older than 0.1\,Gyr
according to the algorithm of \citet{GirBer98} and the mass-loss rate
of \citet{Vassiliadis93}; (iii) four chemical compositions are
considered as listed in Table~\ref{enh-deg}.

\subsection{Library of stellar spectra}\label{lib_spectra}

The library of stellar spectra is taken from \citet{Girardi02}. It
covers a large range of the $\log{\Teff}$ -- $\log{\rm g}$ -- and
[M/H] space. No details are given here. Suffice to mention

\begin{itemize}
\item The basic spectra are from Kurucz ATLAS9 non-overshooting models
\citep{Castelli97,Bessell98} complemented with:

\item Blackbody spectra for \Teff$>$50,000\,K;

\item \citet{Fluks94} empirical M-giant spectra, extended with synthetic
ones in the IR and UV, and modified shortward of 4000\AA\ so as to
produce reasonable \Teff-$(U-B)$ and \Teff-$(B-V)$ relations for
cool giants;

\item \citet{Allard00} DUSTY99 synthetic spectra for M, L and T dwarfs.
\end{itemize}

\noindent
The theoretical broad-band colors used in this study are in the
Johnson-Cousins-Glass UBVRIJHK system, using filter response curves
from \citet{Bessell88} and \citet{Bessell90}.

\begin{table*}
\normalsize
\begin{center}
\caption[]{Abundance ratios for the solar-scaled and \alfe\ mixtures adopted in this study.
The abundances are the same as in \citet{Tantalo04a} but for Ti for which lower
[Ti/Fe] ratios have been adopted: either [Ti/Fe]=0 (case a) or [Ti/Fe]=0.20 (case b).
See the text for details).}
\label{tab-enh}
\small
\begin{tabular*}{110mm}{|c|  c|  c c c|   c c c| c|}
\hline
\multicolumn{1}{|c|}{} &
\multicolumn{1}{c|}{$\Gamma=0$} &
\multicolumn{3}{c|}{$\Gamma=0.35$} &
\multicolumn{3}{c|}{$\Gamma=0.50$}&
\multicolumn{1}{c|}{}\\
\hline
\multicolumn{1}{|c|}{Element} &
\multicolumn{1}{c|}{$A_{el}$} &
\multicolumn{1}{c}{$A_{el}$} &
\multicolumn{1}{c}{$[{X_{el}\over Fe}]$} &
\multicolumn{1}{c|}{$[{X_{el}\over H}]$} &
\multicolumn{1}{c}{$A_{el}$} &
\multicolumn{1}{c}{$[{X_{el}\over Fe}]$} &
\multicolumn{1}{c|}{$[{X_{el}\over H}]$} &
\multicolumn{1}{c|}{Note} \\
\hline
 $O $    & 8.87  & 9.37 & 0.50 &  +0.1434 &  9.57  & 0.70 &  +0.2016& \\
 $Ne$    & 8.08  & 8.37 & 0.29 & --0.0666 &  8.49  & 0.41 & --0.0936& \\
 $Mg$    & 7.58  & 7.98 & 0.40 &  +0.0434 &  8.14  & 0.56 &  +0.0610& \\
 $Si$    & 7.55  & 7.85 & 0.30 & --0.0566 &  7.97  & 0.42 & --0.0796& \\
 $S $    & 7.21  & 7.54 & 0.33 & --0.0266 &  7.67  & 0.46 & --0.0374& \\
 $Ca$    & 6.36  & 6.86 & 0.50 &  +0.1434 &  7.06  & 0.70 &  +0.2016& \\
 $Ti$    & 5.02  & 5.65 & 0.00 & --0.3566 &  5.89  & 0.00 & --0.5013& case a \\
 $Ti$    & 5.02  & 5.65 & 0.20 & --0.1566 &  5.89  & 0.20 & --0.2201& case b \\
 $Ni$    & 6.25  & 6.27 & 0.02 & --0.3366 &  6.28  & 0.03 & --0.4731& \\
 $C $    & 8.55  & 8.55 & 0.00 & --0.3566 &  8.55  & 0.00 & --0.5013& \\
 $N $    & 7.97  & 7.97 & 0.00 & --0.3566 &  7.97  & 0.00 & --0.5013& \\
 $Na$    & 6.33  & 6.33 & 0.00 & --0.3566 &  6.33  & 0.00 & --0.5013& \\
 $Cr$    & 5.67  & 5.67 & 0.00 & --0.3566 &  5.67  & 0.00 & --0.5013& \\
 $Fe$    & 7.50  & 7.50 & 0.00 & --0.3566 &  7.50  & 0.00 & --0.5013& \\
\hline
\end{tabular*}
\end{center}
\end{table*}

\section{Line absorption indices for SSPs}\label{ind_ssp}

In this study we essentially adopt the large grids of line absorption
indices recently calculated by \citet{Tantalo04a} for SSPs over large
ranges of chemical abundances, enhancement factors $\Gamma$, and
ages. However we have slightly changed the ratio $[X_{el}/Fe]$ for some
specific elements. More precisely, the abundance ratios for
solar-scaled and $\alpha$-enhanced mixtures with $\Gamma$=0.35 and
$\Gamma$=0.50 are taken from \citet{Tantalo04a} but for Ti for which
lower [Ti/Fe] are used. They are listed in Table~\ref{tab-enh}, in
which Columns (2), (3) and (6) show the abundance $A_{el}$ of elements
in logarithmic scale, columns (4) and (7) list the ratio
$[X_{el}/Fe]$, and finally columns (5) and (8) show the ratio
$[X_{el}/H]$\footnote{The complete grids of stellar tracks, isochrones, magnitudes, are
available on the web site {\it http://pleiadi.pd.astro.it}, whereas
SSPs colors, line absorption indices can be found on the web site {\it
http://dipastro.pd.astro.it/galadriel}.}.

First of all and in view of the discussion below it is worth
reminding the reader the definition of three indices that are
commonly used but which do not belong to the original Lick system.
They are \MFe, \MgFe\ and \MgFei

\begin{eqnarray}
\MFe    & = & 0.5\times (Fe5270+ Fe5335)  \nonumber
\end{eqnarray}

\begin{eqnarray}
\MgFe        & = & \sqrt{\mgb \times (0.5 \times Fe5270+0.5 \times Fe5335)} \nonumber
\end{eqnarray}

\noindent
and

\begin{eqnarray}
\MgFei       & = & \sqrt{ \mgb \times (0.72\times Fe5270+0.28 \times Fe5335)}  \nonumber
\end{eqnarray}

\begin{figure}
\psfig{file=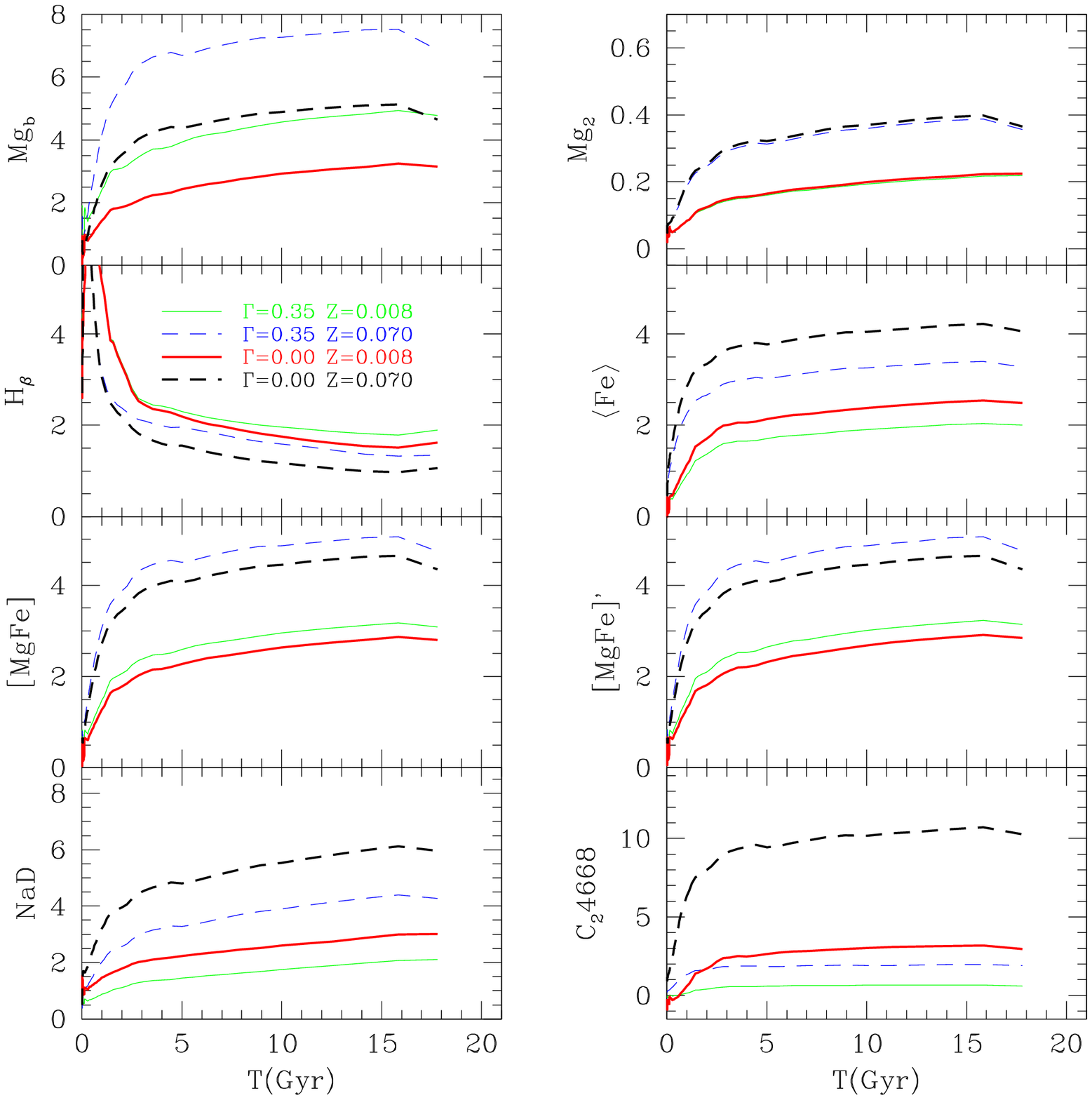,width=8.0truecm}
   \caption{Evolution of eight indices (\mgb, \mgii, \Hbeta, \MFe,
   \MgFe, \MgFei, \nad\ and \cii) as function of the age. The heavy
   lines show the indices for solar-scaled partition of elements
   ($\Gamma$=0), whereas the thin lines show the same but for the
   partition of \alfe\ taken from \citet{Salasnich20}
   ($\Gamma$=0.3557). Only two metallicities are displayed for the
   sake of clarity, i.e. $Z$=0.008 (solid lines) and $Z$=0.07 (dashed
   lines).}
\label{zheh}
\end{figure}

\noindent
Going into a detailed description of the dependence of the line
absorption indices for SSPs on age, chemical composition, $\Gamma$, is
beyond the scope of this study. The reader is referred to
\citet{Tantalo04a} who have discussed these matters in great
detail. Suffice it to show in Fig.~\ref{zheh} the temporal evolution
of eight important indices, i.e. \mgb, \mgii, \Hbeta, \MFe, \MgFe,
\MgFei, \nad\ and \cii, for the following combinations of metallicity and
$\Gamma$, namely $Z$=0.008 (solid lines) and $Z$=0.070 (broken lines),
$\Gamma$=0. (heavy lines) and $\Gamma=0.35$ (light lines). The age
goes from 0.01\,Gyr to 20\,Gyr. It is worth noticing that the index
\mgii\ does not depend on $\Gamma$.

\begin{figure}
\centerline{
\psfig{file=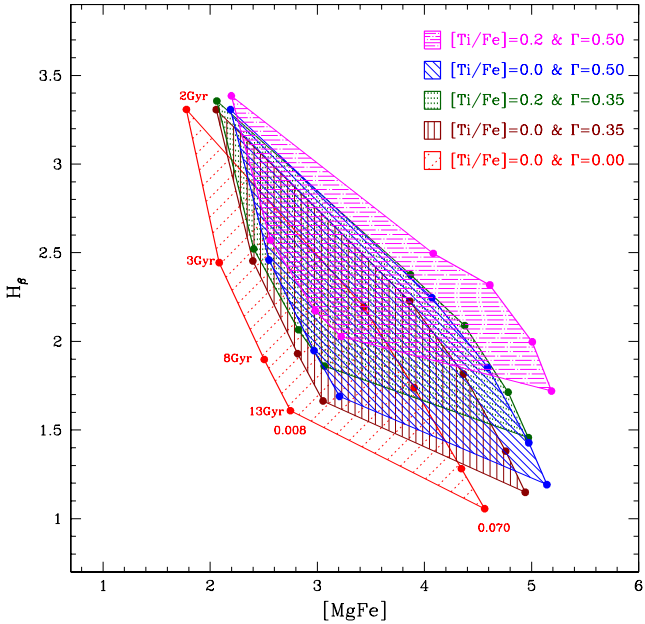,width=9.0truecm}}
   \caption{The \Hbeta\ vs. \MgFe\ plane for different combinations of
   $\Gamma$=0, metallicity, and [Ti/Fe] as indicated. Each hatched
   area is enclosed between the two SSPs with the lowest and highest
   metallicity in our sample, i.e. $Z$=0.008 (left) and $Z$=0.07
   (right). Along each SSP four values of the age are marked so that
   the lines of constant age can be drawn. Four values of the age are,
   i.e. 2, 3, 8, and 14\,Gyr. The different hatched areas correspond
   to different combinations of $\Gamma$ and [Ti/Fe] as indicated in
   the legend. Looking at this diagram one can easily single out the
   separate effect of the four parameters: age, metallicity, $\Gamma$
   and $[X_{el}/Fe]$ ([Ti/Fe] in this case). See the text for more details.}
\label{enh_plane}
\end{figure}

Perhaps the most relevant result here is the variation of the
indices with $\Gamma$. The easiest way of showing this is the
\Hbeta\ vs. \MgFe\ plane shown in Fig.~\ref{enh_plane}, in which
the range of values expected at varying metallicity (heavy solid
lines), age (dotted lines), $\Gamma$ and [Ti/Fe] (hatched areas) as
displayed. Each hatched area is enclosed between the two SSPs with the
lowest and highest metallicity in our sample, i.e. $Z$=0.008 (left)
and $Z$=0.07 (right) and two lines of constant age (14\,Gyr bottom and
2\,Gyr top). Along each SSP four values of the age are marked, so that
other lines of constant age can be drawn. Let us consider the hatched
areas corresponding to $\Gamma$=0, 0.35 and 0.5, independently of the
value for [Ti/Fe], either 0 or 0.2. These cases will be discussed
separately below.  At increasing $\Gamma$ both \Hbeta\ and \MgFe\
increase. While the effect on \MgFe\ in somehow expected, the one on
\Hbeta\ is more difficult to understand. The explanation is in the
strong Response Functions for elements like N, O, Mg etc found by
\citet{Tripicco95} and in the different way enhancing those
elements affects the ratio $F_{l}/F_{c}$. In presence of enhancement,
the spectrum over the three pass-bands defining the index \Hbeta is
more absorbed. However, absorption in the blue wing is larger than in
the central band and in the red wing. Owing to this, the ratio
$F_{l}/F_{c}$ gets smaller so that \Hbeta\ gets stronger. This trend
is also confirmed by the new Response Functions calculated by
\citet{Tantalo04c} using high resolution spectra (1\AA).

There is another important point to be addressed, which has
already been touched upon by \citet{Tantalo04a} and it is shown in
Fig.~\ref{enh_plane} by the cases with the same $\Gamma$, either
0.35 or 0.5, and different [Ti/Fe]. In brief, at given metallicity
$Z$, total enhancement factor $\Gamma$, and list of
enhanced/depressed elements, the same $\Gamma$ can be obtained by
many patterns of $[X_{el}/Fe]$. For instance \citet{Tantalo04a}
have shown that at given $\Gamma$ indices like \Hbeta\ are very
sensitive to the abundance ratio [Ti/Fe]. In their calculations
several values of [Ti/Fe] have been explored, i.e. [Ti/Fe]=0, 0.20
and 0.63. The first choice implies that Ti is not considered as an
$\alpha$-element, the second one is based on the mean value
measured by \citet{Gratton03} for galactic metal-poor stars with
accurate parallaxes, whereas the last choice comes from the old
estimate by \citet{Ryan91} for stars of the same type. In
Fig.~\ref{enh_plane} we show the effect of varying [Ti/Fe] from 0
to 0.2. Test calculations show that other elements like O, Mg, Ne,
Ca etc. have a similar effect. The explanation is the same as
before.

Since there is no unique pattern of abundance ratios as clearly
shown by the above mentioned observational data, at given $Z$ and
$\Gamma$ a sort of natural width for the indices has to be
expected. The natural width we are talking about is provided by
the difference $\left(I_{ [X_{el}/Fe] } - I_{ [X_{el}/Fe]'
}\right)_{age, \Gamma, Z}$. It is soon evident that models with
the same $Z$, $\Gamma$ and age may have significantly different
\Hbeta\ and \MgFe\ (indices in general) depending of the detailed
pattern of [$X_{el}$/Fe] in use.

\section{Composing SSPs}\label{composing}

Since the stellar content of a galaxy can be approximated by a
manifold of SSPs of different age and metallicity, each of which
weighed on the star formation rate, the indices of composite stellar
populations can be derived in simple fashion without making use of
sophisticated population synthesis techniques requiring
chemo-spectro-photometric models of galaxies \citep{Tantalo98b}.

Given the generic index $I_{l}^{SSP}(Y,Z,\Gamma,t)$, with the aid
of eqn.~(\ref{int_ssp}) we derive the integrated fluxes of the SSP
in the absorption feature and pseudo-continuum

\begin{equation}
\mathcal{F}_{l}(Y,Z,\Gamma,t)=\sum_{i} F^{*}_{l,i} N_{i}
\end{equation}

\noindent
and

\begin{equation}
    \mathcal{F}_{c}(Y,Z,\Gamma,t)=\sum_{i} F^{*}_{c,i} N_{i}
\end{equation}

Let us consider a composite object made of several SSPs with assigned
composition, total enhancement and age -- $(Y_{j},Z_{j},\Gamma_{j},t_{j})$ --
each of which intervening in suitable percentages $\beta_{j}$. This
parameter measures the intensity of star formation assigned to each
component, in other words it is the fraction of the galaxy mass
engaged in each SSP. The integrated fluxes are

\begin{equation}
\mathcal{F}_{l}^{comp} = \sum_{j} \beta_{j} \mathcal{F}_{l,j}(Y_{j},Z_{j},\Gamma_{j},t_{j})
\end{equation}

\noindent
and

\begin{equation}
\mathcal{F}_{c}^{comp} = \sum_{j} \beta_{j} \mathcal{F}_{c,j}(Y_{j},Z_{j},\Gamma_{j},t_{j})
\end{equation}

\noindent The total mass is the sum of the SSP mass over all the
components and re-normalization of the fluxes can be applied if
required. The composite indices are then derived by inserting the
above integrated fluxes into eqn.~(\ref{int_ssp}).

\begin{figure}
\centerline{
\psfig{file=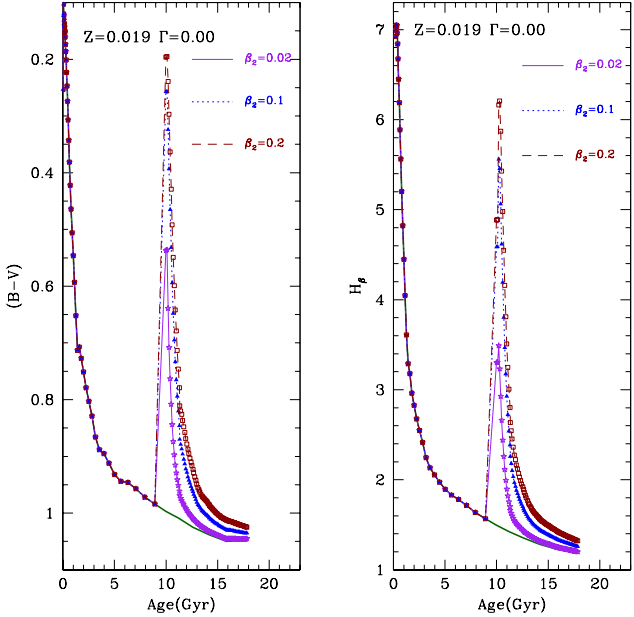,width=9.0truecm}}
   \caption{A burst of star formation of different intensity (as
   indicated) is let occur in an ideal galaxy represented by a SSP
   with $Z$=0.019 and $\Gamma$=0 at the age of 10\,Gyr. The composite
   object is then followed up to the age of 16\,Gyr. {\bf Left Panel}:
   (B--V) vs. age. {\bf Right Panel}: \Hbeta\ vs. age.}
\label{simul_age}
\end{figure}

\begin{figure}
\centerline{
\psfig{file=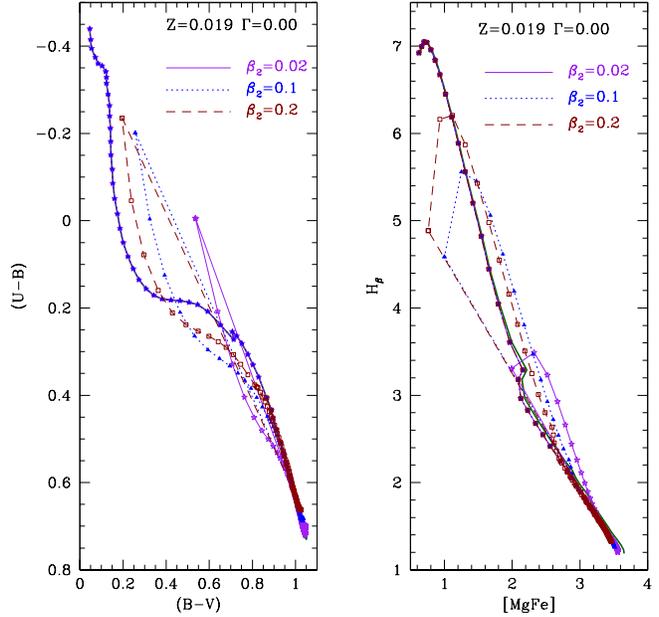,width=9.0truecm}}
   \caption{The same as in Fig.~\ref{simul_age}. {\bf Left panel}: the
   (U--B) vs. (B--V) plane. {\bf Right Panel}: the \Hbeta\ vs. \MgFe\
   plane.}
\label{simul_plane}
\end{figure}

For the sake of illustration we consider the case of an old
galaxy, represented by a SSP with solar composition and no
enhancement of $\alpha$-elements ($Z$=0.019 and $\Gamma$=0), which
at the age of 10\,Gyr suffers a burst of stellar activity
represented by another SSP of the same $Z$ and $\Gamma$. The old
galaxy is labelled by $j$=1 and the burst component by $j$=2.
Three cases of the burst intensity are examined, i.e.
$\beta_{1}$=0.98 and $\beta_{2}$=0.02, $\beta_{1}$=0.90 and
$\beta_{2}$=0.10, $\beta_{1}$=0.80 and $\beta_{2}$=0.20. Once the
burst has occurred the photometric properties of the composite
object are described with the age step of the young SSP in order
not to loose in the time resolution of magnitudes, colors, and
indices. In simulations of this type, both the old and the young
component are let evolve with time from the burst epoch to the
present. The results are shown in the panels of
Fig.~\ref{simul_age} and Fig.~\ref{simul_plane}. The left panel of
Fig.~\ref{simul_age} is the color (B--V) vs. age (in Gyr); whereas
the right panel is the \Hbeta\ vs. age. The left panel of
Fig.~\ref{simul_plane} is the plane (U--B) vs. (B--V), whereas the
right panel is the \Hbeta\ vs. \MgFe\ plane. For the sake of
clarity, the burst is displayed only for ages older than 0.1\,Gyr
because its path in the two colors and/or two indices planes for
younger ages is quite complicated and difficult to describe in a
simple fashion. In the left and right panels of
Fig.~\ref{simul_age} the age structure of the bursting mode is
clear. The color (B--V) and index \Hbeta\ are suddenly rejuvenated
at the onset of the burst: (i) (B--V) may get very blue, the peak
value depends on $\beta_{2}$, and then fade down toward the
typical color of an old object as the age increase from 10\,Gyr up
to 16\,Gyr. The same for \Hbeta. (ii) In all simulations both
\Hbeta\ and (B--V) keep memory of the bursting activity for long
time. Both indeed tend to be lower and higher, respectively, than
the pure passive evolution indicated by the solid heavy line. In
the planes (U--B) vs. (B--V) and \Hbeta\ vs. \MgFe\ of
Fig.~\ref{simul_plane}, the composite object performs wide loops,
the extension and width of which are functions of the intensity
$\beta_{2}$. The complete path of the bursting object is better
shown in Fig.~\ref{comp_ind} below.

\begin{figure}
\centerline{
\psfig{file=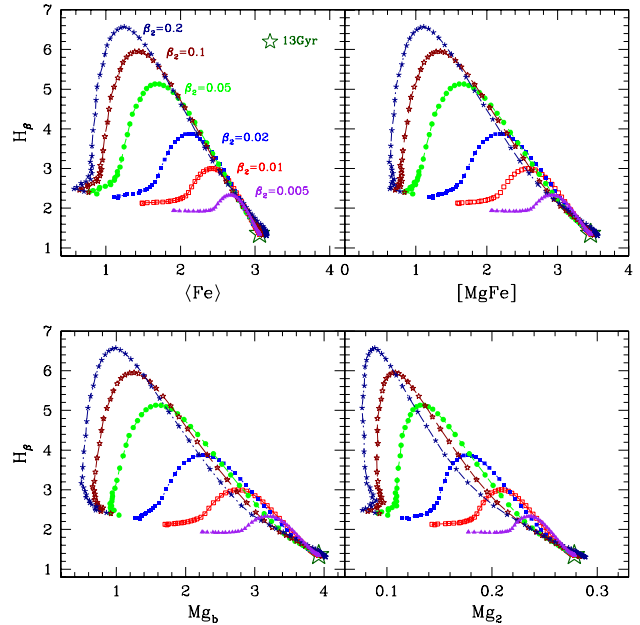,width=9.0truecm}}
   \caption{Composite SSPs: young SSPs of any age and metallicity
   $Z$=0.040 are superposed to an old SSP with the age of 13\,Gyr and
   the metallicity $Z$=0.019. The age of the old SSP is kept
   constant. Different percentages ($\beta_{2}$) for the young
   component are considered as indicated. The percentage of the old
   component is $\beta_{1}$=$1 - \beta_{2}$. The various panels show
   four different diagnostic planes. }
\label{comp_ind}
\end{figure}

Another interesting plane to look at is the superposition of two
SSPs, one old and the other young (in different percentages as
above, but in which the age of the old SSP is kept fixed whereas
that of the young one is let vary from 0.01\,Gyr to the age of the
old component in suitable steps. These simulations allow us to
span the whole range of possible combinations of ages for the two
SSPs and the whole range of values in any two-indices plane at
varying intensity and age of the burst. In other words, any point
in these plane is a picture of the composite galaxy taken at a
certain age of the burst. Simulations of this type are shown in
Fig.~\ref{comp_ind} for the typical age of the old SSP of 13\,Gyr.
The apparently strange behavior of the SSP path at increasing age
of the burst deserves some explanation. The bell-shaped trend of
\Hbeta\ has already been explained by \citet{Buzzoni94}. In brief for 
strong \Hbeta\ absorption (as in young SSPs and/or burst of star
formation), the Stark wings of the feature overflow the Lick wavelength
window, and enter the side bands depressing the pseudo-continuum.
As a result, the \Hbeta\ index peaks when the SSP is dominated by B5-A0
stars and fades for earlier and later spectral types. 
This is mirrored in the path of the composite SSP with the contribution 
from each component weighed on its percentage. It is worth noticing that 
a galaxy caught at the very early stages of its bursting activity would
appear as an object with unusually low indices. We will come back
to this later on.

\begin{figure*}
\centerline{
\psfig{file=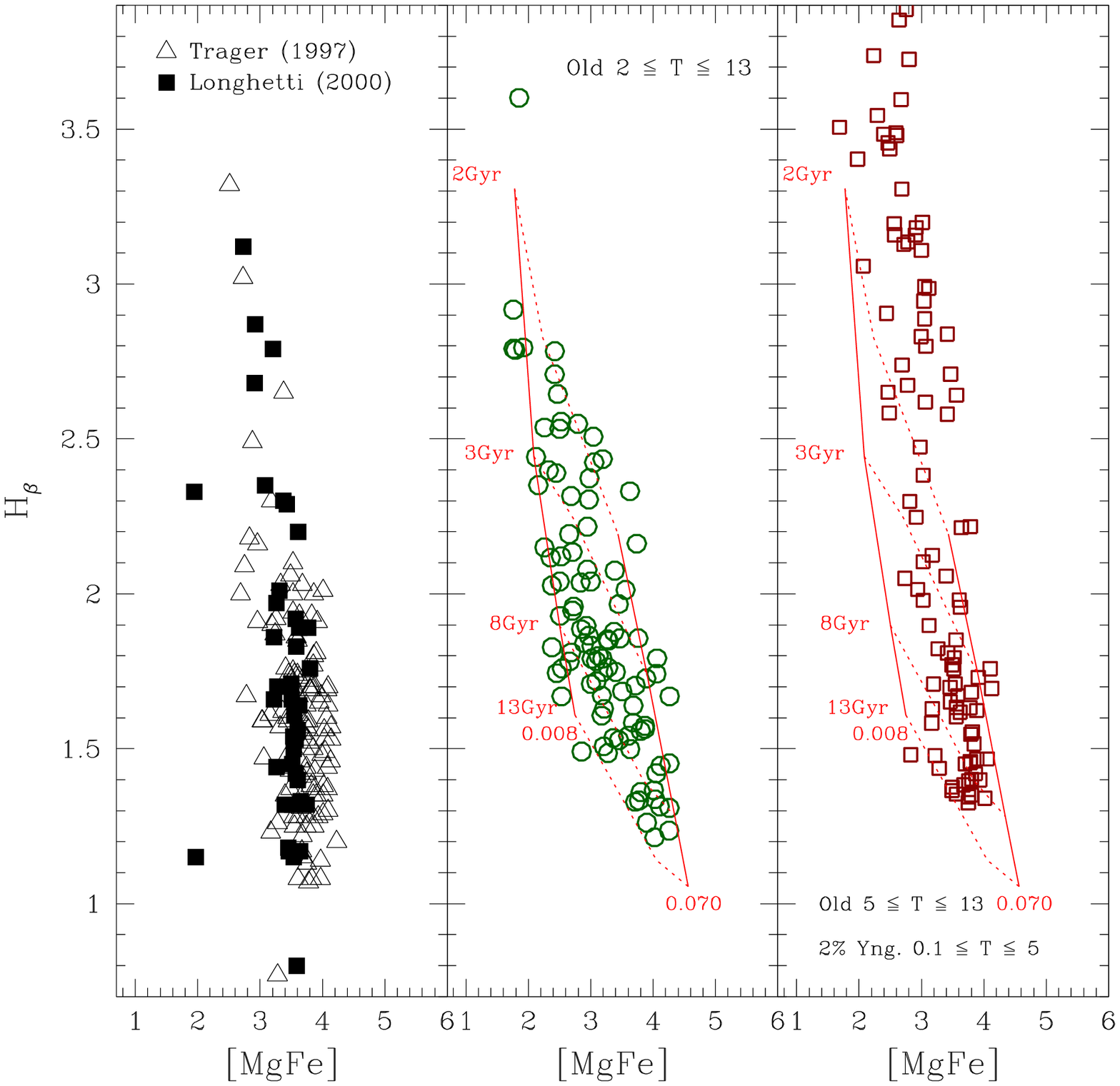,width=15.0truecm,height=9.0truecm}}
   \caption{Observational data and MonteCarlo simulations in the
   \Hbeta\ vs. \MgFe\ plane. {\bf Left Panel}: The normal galaxies by
   \citet{Trager97} indicated by the open triangles and the shell- and
   pair-galaxies of \citet{Longhetti20} indicated by the filled
   squares. Note the smooth and broad distribution of both groups of
   galaxy. {\bf Central Panel}: MonteCarlo simulations for the set of
   parameters $T_{1}$=13\,Gyr, $T_{2}$=2\,Gyr, $\beta_{1}$=1, no
   additional bursts of star formation.  The metallicity is randomly
   chosen in the interval 0.008$\leq$$Z$$\leq$0.07, no enhancement of
   $\alpha$-element is considered ($\Gamma$=0). The solid lines are
   two SSPs with $Z$=0.008 (left) and $Z$=0.07 (right). Four values of
   the age, i.e. 2, 3, 8 and 13\,Gyr are marked. The thin dotted lines
   are the loci of constant age. The very large age range is required
   to get full coverage of the range in \Hbeta\ spanned by the
   data. {\bf Right Panel}: MonteCarlo simulations according to the
   recipe: $T_{1}$=13\,Gyr, $T_{2}$=5\,Gyr, random burst of star
   formation activity at the time 0.1$\leq T_{3} \leq$10, the same
   interval for the minimum and maximum metallicity and the same
   $\Gamma$ as in the previous simulations. The percentage of the
   young population is $\beta_{2}$=0.02. The bulk population formed in
   the time interval $T_{1}$ to $T_{2}$ obeys the universal law of
   metal enrichment. See the text for details. All other symbols have
   the same meaning as in the Central Panel.}
\label{hb_mgfe_simul}
\end{figure*}

\section{Index-index diagnostics: age or chemistry?}\label{sim_ssp_age}

The sample of galaxies we intend to analyze is derived from two
sources of data, namely  the ``{\it IDS Pristine}'' catalog by
\citet{Trager97} for the central regions of normal galaxies -- it
contains also the \citet{Gonzalez93} list --, and the catalog by
\citet{Longhetti20} for pair- and shell-galaxies (i.e. only
objects with clear signs of interaction). The latter sample is
given in Table~\ref{longhetti}\footnote{We remind the reader that
the values of \MgFe\ listed in Table~4 of \citet{Longhetti20} have
been calculated using the wrong expression \MgFe$= \sqrt{\mgb
\times Fe5270+Fe5335/2}$}. The \Hbeta\ vs.\MgFe\ plane for all the
galaxies in question is shown in the left panels of both
Figs.~\ref{hb_mgfe_simul} and Fig.~\ref{enh_simul} to facilitate
the comparison with the theoretical simulations we are going to
present. It is worth calling attention on the striking similarity
between normal and interacting galaxies, and the very smooth but steeper 
slope of the data as compared to that  of SSPs of given metallicity 
(see Fig. \ref{enh_plane}). In the following we will refer to it
as the {\it nearly vertical distribution} along the \Hbeta\ axis.

\subsection{A toy galaxy model}\label{toy_model}

\citet{Bressan96} and \citet{Longhetti20} suggested that the distribution
of galaxies in the \Hbeta\ vs. \MgFe\ plane may reflect secondary
episodes of star formation superposed to an old stellar component. In
the following, we intend to explore further this idea by means of
MonteCarlo simulations based on the \citet{Longhetti20} toy model of
galaxy evolution.

The complex star formation history of an early-type galaxy is
reduced to a burst of relatively recent star formation superposed
to the bulk population made of old stars. These latter are in turn
represented by a SSP whose age is randomly selected between the
ages $T_{1}$ and $T_{2}$. The young stellar component, formed
during the recent burst of star formation, is represented by a SSP
whose age is randomly selected between $T_{2}$ and a lower limit
$T_{3}$. Typical values for the age limits are: $T_{1}$=13,
$T_{2}$=10, and $T_{3}$=0.1\,Gyr. The time-scale of the two star
forming event are always assumed to be short compared to all other
relevant timescales (ages, and Hubble time).

Since we are interested in guessing the minimum threshold above which
the secondary episode gets importance in affecting the line strength
indices, we will consider only the case in which the secondary episode
involves a minor fraction of the galaxy mass. Typical values for the
strength of the primary and secondary burst $\beta_{1}$ and
$\beta_{2}$, respectively, measuring the percentage of the total
galaxy mass turned into stars, are $\beta_{1}$=0.98 and
$\beta_{2}$=0.02.

The metallicity $Z_{1}$ of the old stellar component is randomly
selected between 0.008 (50\% of the solar value) and 0.07 (3.25
the solar value). We have also built a set of simulations in which
the metallicity of the old component is forced to linearly
increase from 0.008 and 0.07 over the age range $T_{1}$ to
$T_{2}$. The metallicity $Z_{2}$ of the young component is
randomly chosen over the whole range (i.e. from $Z$=0.008 to
$Z$=0.07), thus simulating the widest range of possibilities,
going from acquisition of external less processed gas to chemical
enrichment during the burst.

The simulations are first performed for solar partitions of elements,
i.e. $\Gamma$=0 and then for $\Gamma\neq$0 (another dimension is added
to the problem).

Finally, random errors are applied to the model indices to better
simulate the observations. Using the data by \citet{Trager97} we
calculate the mean relative errors $\langle \Delta I/I
\rangle_{0}$ as reported in Table~\ref{errors}. The error
affecting an index is randomly evaluated according to

\begin{equation}
\Delta I= - \left\langle \frac{\Delta I}{I} \right\rangle_{0} I +
2\times \epsilon\,
            \left\langle \frac{\Delta I}{I} \right\rangle_{0} I
\end{equation}

\noindent
where $\epsilon$ is a random number between 0 and 1.

\subsection{Large scatter in age and metallicity of the bulk population}\label{age_spread}

This is the simplest interpretation of the distribution of the
galaxies in the \Hbeta\ vs. \MgFe\ plane. Neglecting important
effects due to enhancement in $\alpha$-elements and recent stellar
activity, matching the observational range of the data would
require the bulk population of early type galaxies being formed in
different epochs from galaxy to galaxy over a time-scale
comparable to the Hubble time. In the central  panel of
Fig.~\ref{hb_mgfe_simul} we show a simulated sample of 100 objects
for which the following parameters are assumed: $T_{1}$=13\,Gyr,
$T_{2}$=2\,Gyr, $\beta_{1}$=1, no increase of the metallicity in
this time interval, no later bursts of stellar activity, and
finally $\Gamma$=0. The models galaxies distribute along the SSP
lines of different metallicity according to their age. The
simulation significantly differs from the observational data shown
in left panel of Fig.~\ref{hb_mgfe_simul}. Even if this view could
be fitted into the classical hierarchical scheme of galaxy
formation, it can be hardly sustained because it would predict
spectro-photometric properties not fully compatible with the
observational data for EGs.

\begin{table*}
\caption[]{Basic data for the \citet{Longhetti20} sample of shell-
and pair-galaxies. The velocity dispersion $\Sigma_{0}$ is km/s.}
\begin{scriptsize}
\begin{center}
\begin{tabular*}{165mm}{|l r r r r r r r| l r r r r r r r|}
\hline
\multicolumn{8}{|c}{Pair-Galaxies}&\multicolumn{8}{|c|}{Shell-Galaxies}\\
\hline
\multicolumn{1}{|l}{Name} &
\multicolumn{1}{c}{\Hbeta} &
\multicolumn{1}{c}{\mgii} &
\multicolumn{1}{c}{\mgb} &
\multicolumn{1}{c}{Fe52} &
\multicolumn{1}{c}{Fe53} &
\multicolumn{1}{c}{\MgFe} &
\multicolumn{1}{c|}{$\Sigma_{0}$} &
\multicolumn{1}{l}{Name} &
\multicolumn{1}{c}{\Hbeta} &
\multicolumn{1}{c}{\mgii} &
\multicolumn{1}{c}{\mgb} &
\multicolumn{1}{c}{Fe52} &
\multicolumn{1}{c}{Fe53} &
\multicolumn{1}{c}{\MgFe} &
\multicolumn{1}{c|}{$\Sigma_{0}$} \\
\hline
 RR24a    & -2.60 & 0.17 & 2.49 & 2.22 & 2.32 & 2.38 & 101 &
 N813     &  2.29 & 0.25 & 4.19 & 3.33 & 2.28 & 3.43 & 208 \\
 RR24b    &-17.37 & 0.09 & 2.01 & 0.74 & 1.10 & 1.36 &  92 &
 N1210    &  1.44 & 0.26 & 3.94 & 2.92 & 2.46 & 3.26 & 180 \\
 RR62a    &  1.15 & 0.13 & 2.30 & 1.57 & 1.82 & 1.97 &  81 &
 N1316    &  2.01 & 0.26 & 4.02 & 2.84 & 2.62 & 3.31 & 250 \\
 RR101a   &  1.86 & 0.23 & 3.53 & 3.10 & 2.76 & 3.22 & 162 &
 N1549    &  1.71 & 0.29 & 4.46 & 2.85 & 2.64 & 3.50 & 225 \\
 RR101b   &  2.35 & 0.23 & 3.67 & 2.83 & 2.35 & 3.08 & 210 &
 N1553    &  1.40 & 0.30 & 4.45 & 3.05 & 2.79 & 3.60 & 188 \\
 RR105a   &  1.32 & 0.27 & 4.23 & 3.21 & 2.26 & 3.40 & 130 &
 N1571    &  1.66 & 0.28 & 4.31 & 3.24 & 2.48 & 3.51 & 224 \\
 RR187b   &  2.87 & 0.22 & 3.40 & 2.69 & 2.34 & 2.92 & 144 &
 N2865    &  3.12 & 0.22 & 3.28 & 2.34 & 2.22 & 2.22 & 208 \\
 RR210a   &  0.80 & 0.33 & 4.97 & 2.71 & 2.47 & 3.59 & 281 &
 N2945    &  0.38 & 0.26 & 4.80 & 2.62 & 2.21 & 3.40 & 212 \\
 RR210b   &  1.50 & 0.31 & 4.62 & 2.99 & 2.44 & 3.54 & 212 &
 N3051    &  1.15 & 0.30 & 4.82 & 2.74 & 2.45 & 3.54 & 212 \\
 RR225a   &  1.76 & 0.34 & 4.96 & 3.41 & 2.40 & 3.80 & 357 &
 N5018    &  2.68 & 0.22 & 3.31 & 2.89 & 2.24 & 2.91 & 247 \\
 RR225b   &  1.18 & 0.31 & 4.42 & 2.80 & 2.59 & 3.45 & 186 &
 N6776    &  1.92 & 0.25 & 4.39 & 2.67 & 3.15 & 3.57 & 243 \\
 RR278a   & -1.67 & 0.20 & 4.14 & 2.20 & 2.44 & 3.10 & 184 &
 N6849    &  1.32 & 0.28 & 4.34 & 3.03 & 2.67 & 3.52 & 198 \\
 RR282b   &  1.53 & 0.33 & 5.17 & 2.78 & 2.16 & 3.57 & 290 &
 N6958    &  1.66 & 0.26 & 4.08 & 2.80 & 2.29 & 3.22 & 223 \\
 RR287a   &  1.61 & 0.28 & 4.57 & 3.08 & 2.43 & 3.55 & 205 &
 E1070040 &  2.30 & 0.25 & 3.97 & 2.98 & 2.74 & 3.37 & 155 \\
 RR297a   &  1.97 & 0.27 & 4.08 & 2.89 & 2.32 & 3.26 & 177 &
 E3420390 &  1.32 & 0.33 & 4.77 & 3.15 & 2.75 & 3.75 & 327 \\
 RR297b   &  1.17 & 0.26 & 4.81 & 2.92 & 2.60 & 3.64 & 147 &
 N7135    & -0.41 & 0.31 & 5.51 & 2.81 & 2.38 & 2.38 & 191 \\
 RR298b   &  1.42 & 0.31 & 4.48 & 3.24 & 2.44 & 3.57 & 130 &
 E2890150 &  0.29 & 0.20 & 3.06 & 2.22 & 2.47 & 2.68 & 274 \\
 RR307a   &  0.08 & 0.21 & 3.73 & 1.59 & 1.53 & 2.41 &  92 &
 E2400100a&  1.54 & 0.28 & 4.63 & 2.66 & 2.71 & 3.53 & 225 \\
 RR317a   &  1.56 & 0.28 & 4.36 & 3.06 & 2.90 & 3.60 & 189 &
 E2400100b&  2.79 & 0.21 & 3.79 & 3.10 & 2.33 & 3.21 & 223 \\
 RR317b   &  2.33 & 0.13 & 1.80 & 2.47 & 1.74 & 1.95 &  91 &
 E5380100 &  1.70 & 0.31 & 4.20 & 3.01 & 2.10 & 3.28 & 178 \\
 RR381a   &  1.64 & 0.31 & 5.01 & 2.63 & 2.62 & 3.63 & 276 &
          &       &      &      &      &      &      &     \\
 RR387a   &  2.20 & 0.29 & 4.48 & 3.08 & 2.74 & 3.61 & 187 &
          &       &      &      &      &      &      &     \\
 RR387b   &  1.83 & 0.31 & 4.51 & 3.13 & 2.55 & 3.58 & 215 &
          &       &      &      &      &      &      &     \\
 RR397b   &  1.33 & 0.29 & 4.39 & 3.21 & 2.87 & 3.65 & 198 &
          &       &      &      &      &      &      &     \\
 RR405a   &  1.45 & 0.28 & 4.68 & 2.92 & 2.39 & 3.52 & 192 &
          &       &      &      &      &      &      &     \\
 RR405b   &  1.89 & 0.34 & 4.81 & 2.72 & 2.76 & 3.63 & 207 &
          &       &      &      &      &      &      &     \\
 RR409a   &  1.89 & 0.29 & 4.51 & 3.02 & 3.28 & 3.77 & 187 &
          &       &      &      &      &      &      &     \\
 RR409b   &  1.17 & 0.31 & 4.56 & 2.89 & 2.36 & 3.46 & 150 &
          &       &      &      &      &      &      &     \\
\hline
\end{tabular*}
\end{center}
\end{scriptsize}
\label{longhetti}
\end{table*}

\subsection{Random bursts of star formation} \label{random_bursts}

Bursts of star formation (from one to several) superposed to an
old dominant population of stars seem to be more plausible and
yield better results. Galaxies are conceived as old, nearly coeval
systems, their population being approximated by a single SSP with
age between 10 and 13\,Gyr; this age range agrees with the current
age estimate of EGs in rich clusters \citep{Bower98}. A burst of
stellar activity is added at an age randomly chosen in the
interval $T_{2}$--$T_{3}$. Two different prescriptions for the
metallicity are adopted as described in Section~\ref{toy_model}
above. Finally all the simulations are for $\Gamma$=0. The maximum
intensity of the superposed burst amounts to 2\% of the total
mass, i.e. $\beta_{1}$=0.98 and $\beta_{2}$=0.02.

The simulations of this type are made at increasing complexity. Since
they essentially confirm what already found by \citet{Longhetti20} we
limit ourselves to discuss the results and to highlight the point of
disagreement with the observational data without showing any
simulation in detail:

(i) Stronger bursts of star formation (i.e. engaging more than 2\% of
the mass) are nor suited as they would predict too high values of
\Hbeta\ and too many {\it young} objects.

(ii) The expected distribution of objects with respect to the
\Hbeta\ index is at variance with the observational one.
Indeed models of this type predict a bimodal distribution, whereby the
old galaxies (those for which the burst is almost as old as the bulk
of their stellar populations) clump together in the lower portion of
the diagram, whereas the ``young'' objects (those with very young
bursts) form a tail extending to high values of \Hbeta.

(iii) Nevertheless, the burst alone cannot explain the smooth
distribution observed at low \Hbeta\ values. This is because the
\Hbeta\ index of a stellar population for which the 2\% of the
mass is composed by ``young'' stars and the remaining 98\% by an
old component, reaches the observed high values, but, fading very
rapidly with the age of the ``young'' component, has low
probability to match the intermediate observed values. Slowing
down the index decrease (corresponding to the aging of the
``young'' component) by increasing the percentage of mass involved
by the young burst produces a uncomfortably large fraction of
objects in the upper part of the diagram. A complex interplay
between burst intensity and mean age of the stellar population
should then take place, with the old bursts being on average
stronger than the recent ones. It must be said, however, that when
the burst itself is larger than a few percent of the total mass,
the definition of the average age of the bulk of the stellar
population becomes a problem.

(iv) The difficulty is partially cured assuming that the {\it old}
population has an average age spreading over a significant
fraction of the Hubble time (say down to $T_{2}\simeq$5\,Gyr).
This is meant to indicate that either the object has been growing
for such a long time with a low star formation rate, or that its
major star formation activity was not confined to an early epoch.
The young component is left to occur. It appears immediately that
the observed smooth distribution in the \Hbeta\ index together
with the young tail can be much better reproduced with such a kind
of models, see \citet{Longhetti20} for details.

(v) Another problem arises if the metallicity is randomly
selected. The distribution of the models galaxies in the left panel of
Fig.~\ref{hb_mgfe_simul} strictly follows the path of a SSP, whereas
the data run much steeper. We take this point to invoke the existence
of a relation between the age of the bulk stellar population and its
average metallicity and to apply the ULME. The simulations better fit
the data.

The right panel of Fig.~\ref{hb_mgfe_simul} shows our final
experiment incorporating all the hints we have been discussing so
far. The simulations are based on the following parameters and
assumptions: $T_{1}$=13\,Gyr, $T_{2}$=5\,Gyr, $T_{3}$=0.1\,Gyr,
$\beta_{1}$=0.98, $\beta_{2}$=0.02, average metallicity of the
bulk of the stellar population forced to linearly increase from
$Z$=0.008 to $Z$=0.070 over the time interval $T_{1}$ to $T_{2}$.
Thanks to the combined effect of the large age and metallicity
spread for the bulk population, the distribution in \Hbeta\ vs.
\MgFe\ plane is {\it nearly vertical}. The larger range of metallicity
is necessary to maintain a significant dispersion in \MgFe,
because age differences tend to compensate metallicity
differences. If the latter interpretation is correct, it suggests
that young early-type galaxies in the field are on average more
metal-rich than old systems, with an average metallicity gradient
of about $\Delta$log(Z)/$\Delta$log(t)$\simeq -$0.7, the latter
value being very dependent on the younger limit of the bulk age.

The main conclusion out of these simulations is that in addition to
the mass dominating old population, which however has to be built up
over a large time interval and under a suitable age-metallicity
relationship, sprinkles of stellar activity in the recent or very
recent past ought to considered in nearly all galaxies to reconcile
theory and observations. {\it However, that all galaxies have to go
through recent star forming activity is perhaps too demanding and
other alternatives should be explored}.

\begin{table*}
\small
\begin{center}
\caption[]{Relative errors for the theoretical simulations.}
\label{errors}
\begin{tabular*}{166mm}{|c c c c c c c|}
\hline
\multicolumn{1}{|c}{\rule{0cm}{0.35cm} $\left\langle \Delta \Hbeta/\Hbeta \right\rangle$} &
\multicolumn{1}{c}{$\left\langle \Delta \mgb/\mgb     \right\rangle$} &
\multicolumn{1}{c}{$\left\langle  \Delta \mgii/\mgii   \right\rangle$} &
\multicolumn{1}{c}{$\left\langle  \Delta \cii/\cii     \right\rangle$} &
\multicolumn{1}{c}{$\left\langle  \Delta \nad/\nad     \right\rangle$} &
\multicolumn{1}{c}{$\left\langle  \Delta \MFe/\MFe     \right\rangle$} &
\multicolumn{1}{c|}{$\left\langle \Delta \MgFe/\MgFe   \right\rangle$} \\
\hline
\rule{0cm}{0.35cm} 0.136 & 0.059 & 0.026 & 0.009 & 0.054 & 0.073 & 0.047 \\
\hline
\end{tabular*}
\end{center}
\end{table*}

\subsection{The $\alpha$-enhancement alternative}\label{alpha_avenue}

The results obtained by \citet{Tantalo04a} for SSPs of the same
age and metallicity but different degrees of enhancement in \alfa\
offer a third plausible explanation. The bottom line of the model
is best explained by comparing the theoretical models in the
\Hbeta\ vs. \MgFe\ plane of Fig.~\ref{hb_mgfe_simul} (central and
right panels), with the observational data (left panel). Several
points are soon evident:

(i) The majority of galaxies (those with \Hbeta$\leq$2) are fully
compatible with being very old objects of the same age (say about
13\,Gyr) but a different degree of enhancement in
$\alpha$-elements going from $\Gamma$=0 to $\Gamma$=0.5 (taking
the so-called natural width caused by a possible variation in
single elemental species into account). As a matter of fact, an
old galaxy (say a 10-13\,Gyr object) being shifted to higher
\Hbeta\ by high $\Gamma$ and/or [$X_{el}$/Fe] ([Ti/Fe] as a
prototype) could lie in the same region occupied by a galaxy of
significantly younger age and solar abundance ratios. At least
part of the scatter along the \Hbeta\ axis could be due to a
different degree of enhancement in \alfa.

(ii) Only for galaxies with \Hbeta$>$2, unless their enhancement
factor $\Gamma$ and abundance ratios [$X_{el}$/Fe] (like [Ti/Fe])
are larger than the above limits, the presence of secondary star
forming activity ought to be invoked.

(iii) Looking at the position of models of constant metallicity
and age but different $\Gamma$ and/or [$X_{el}$/Fe] (for instance
[Ti/Fe]), they scatter along a {\it nearly vertical} line. This implies
that the metallicity relationship invoked for the bulk old
population is no longer required. The {\it vertical} distribution
of the data is simply caused by the compensatory effect of
different combinations of $Z$, $\Gamma$ and  [$X_{el}$/Fe]
([Ti/Fe] in our case).

(iv) Secondary episodes of star formation are no longer a common
feature to all galaxies, but an exceptional event limited to a small
number of them. This agrees with the age distribution obtained by
\citet{Tantalo04a}.

\begin{figure*}
\centerline{
\psfig{file=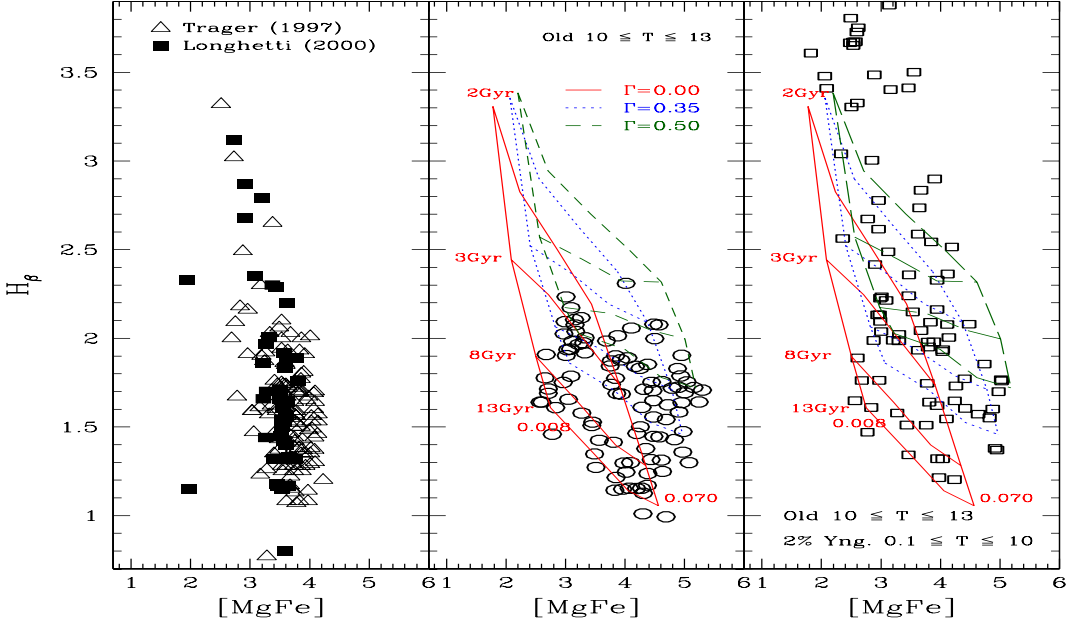,width=15.0truecm,height=9.0truecm}}
   \caption{Observational data and MonteCarlo simulations for the
   $\alpha$-Model. {\bf Left Panel}: The normal galaxies by
   \citet{Trager97} indicated by the open triangles and the shell-
   and pair-galaxies of \citet{Longhetti20} indicated by the filled
   squares. {\bf Central Panel}: old galaxies formed in the time
   interval $T_{1}$=13 to $T_{2}$=10\,Gyr and no later star formation
   activity ($\beta_{1}$=1 and $\beta_{2}$=0). Each model galaxy has
   a metallicity and enhancement factor $\Gamma$ randomly chosen from
   the whole range, i.e. 0.008$\leq$$Z$$\leq$0.07 and 0$\leq$$\Gamma$$\leq$0.35.
   No universal enrichment law is considered. Thanks only to the
   different combinations of $Z$ and $\Gamma$ (different chemical
   compositions in general), the model galaxies span a large range of
   the observational values for \Hbeta\ and \MgFe. {\bf Right panel}:
   the same as before, but now a small percentage of a young
   population has been added, i.e. $\beta_{1}$=0.98 and
   $\beta_{2}$=0.02, whose age randomly varies in the interval
   0.1$\leq$$T_{3}$$\leq$10\,Gyr. The secondary burst of star
   formation has been included to get full coverage of the
   observational \Hbeta\ range in particular for those galaxies with
   \Hbeta$>$2.5. In both panels the meaning of the various lines
   (solid, dashed and dotted) is the same as in Fig.~\ref{enh_plane}.}
\label{enh_simul}
\end{figure*}

The $\alpha$-Model is confirmed by the MonteCarlo simulations
shown in the left and right panels of Fig.~\ref{enh_simul}. In the
left panel we show the case of old galaxies with no secondary
activity: the bulk population spans the age range given by
$T_{1}$=13\,Gyr and $T_{2}$=10\,Gyr with $\beta_{1}$=1, whereas
the metallicity and $\alpha$-enhancement span the whole range for
the parameters $Z$, $\Gamma$, and [$X_{el}$/Fe] ([Ti/Fe] taken as
a measure of the natural width). Since there is no further star
formation activity $T_{3}$=10\,Gyr and $\beta_{2}$=0. The models
essentially match the bulk of data, i.e. galaxies with
\Hbeta$\leq$2, and yield a distribution in the \Hbeta\ vs. \MgFe\
plane which is {\it nearly vertical} (no memory of the SSPs path). In
the right panel we show the same but allowing for recent burst to
occur, i.e. $T_{3}$=0.1\,Gyr. The burst intensity is for
$\beta_{2}$=0.02. But for the few galaxies clearly caught in the
burst mode (those with \Hbeta$>$2), the two theoretical
distributions are nearly identical and both fairly well reproduce
the observational data.

These simulations open the gate to an interesting alternative
explanation, i.e. that the large scatter in \Hbeta\ and \MgFe\ is
predominately caused by a spread in the chemical parameters
metallicity $Z$ and enhancement factor $\Gamma$ rather than
metallicity and age in the bulk population of a galaxy. Secondary
activity of star formation is unavoidable only for a minority of
objects.

The scatter in $Z$, $\Gamma$ and also individual $[X_{el}/Fe]$ of the
dominant old stellar component could be attributed to different kinds
of star formation at the very early epochs, perhaps related to the
physical conditions in the proto-galaxy affecting not only the
intensity and duration of the star formation process, but also the
initial mass function of the stars and the abundance ratios in
turn. We will touch upon this point later.

\begin{figure}
\centerline{
\psfig{file=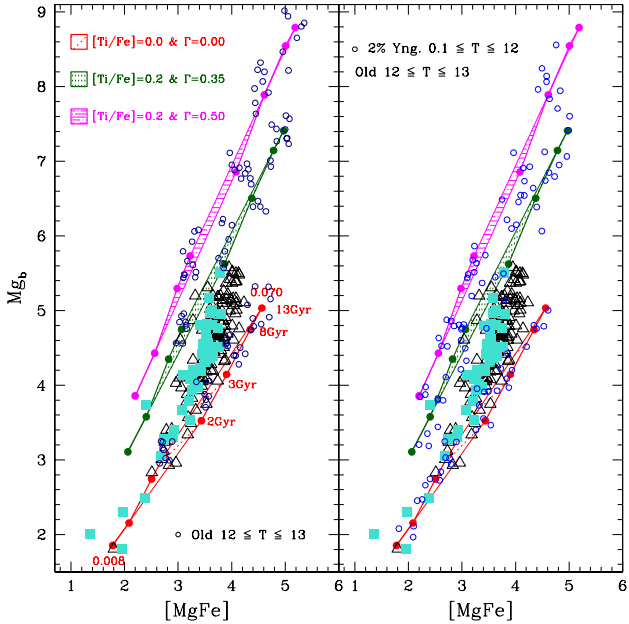,width=9.0truecm}}
   \caption{MonteCarlo simulations of the type shown in
   Fig.~\ref{enh_simul} but with a narrower age range for the old
   component. They are aimed to better constrain the metallicity and
   enhancement factor suited to the data. {\bf Left Panel}: models
   with only the very old component. {\bf Right Panel}: models with
   both old and young stellar populations. In both panels the
   simulations are indicated by open circles and all other symbols
   have the same meaning as in Figs.~\ref{enh_plane} and
   \ref{enh_simul}. In this plane real galaxies seem to fall in the
   ranges 0$\leq$$\Gamma$$\leq$0.3 and 0.008$\leq$$Z$$\leq$0.05.}
\label{Mgb_MgFe_simul}
\end{figure}

\begin{figure}
\centerline{
\psfig{file=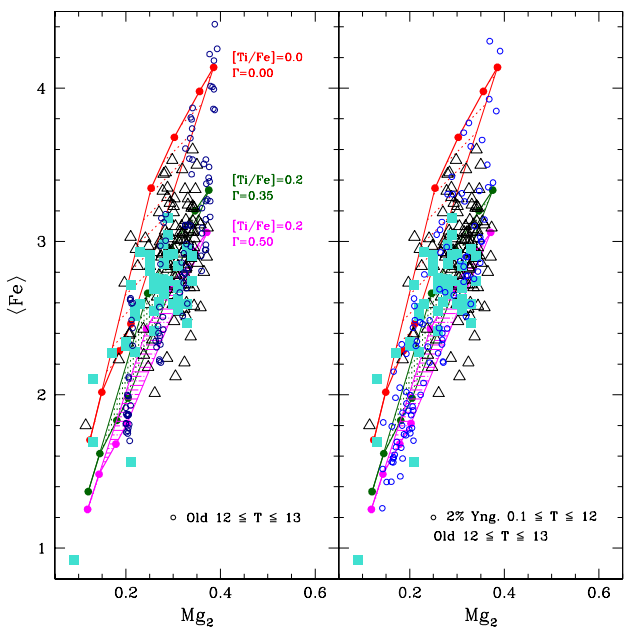,width=9.0truecm}}
   \caption{The same as in Fig.~\ref{Mgb_MgFe_simul} but for the \MFe\
   vs. \mgii\ plane. Now the data seem to be compatible with
   0.35$\leq$$\Gamma$$\leq$0.50 and metallicities extending up to
   $Z$=0.07.}
\label{Fe_Mg2_simul}
\end{figure}

\section{Discussion }\label{discussion}

{\bf Testing the $\alpha$-Model.}
To corroborate the suggestion that the scatter in the \Hbeta\
vs. \MgFe\ plane for most galaxies is due the chemical parameters
$\Gamma$, $Z$, and perhaps [Ti/Fe] we perform here a few {\it ad hoc}
experiments.

First of all, we test a model with a very narrow age range for the
old component and no subsequent star formation. The test is meant
to isolate the sole effect of the chemical abundances. The model
is characterized by the parameters: $T_{1}$=13\,Gyr,
$T_{2}$=12\,Gyr, $\beta_{1}$=1, $T_3$=12\,Gyr, and $\beta_2$=0.
While no ULME is considered, the metallicities, $\Gamma$s and
[$X_{el}$/Fe]s (at present [Ti/Fe]) are let span the whole range.
The simulations are shown in left panel of
Fig.~\ref{Mgb_MgFe_simul} correlating \mgb\ with \MgFe. The choice
of these two indices is based on the notion that \MgFe\ is nearly
independent on $\Gamma$, whereas \mgb\ does. Both are sensitive to
the age and metallicity even if some degree of degeneracy is
present. On purposes we avoided \Hbeta\ because of its equal
sensitivity to all the parameters in question. In this plane we
also displays the theoretical areas corresponding to the different
combinations of $\Gamma$ and [Ti/Fe], metallicities, and ages.
Each area is bounded by the SSPs with the lowest ($Z$=0.008) and
highest ($Z$=0.070) metallicities, the two heavy lines along which
four values of the age are marked (2, 3, 8, and 13\,Gyr). It is
worth noticing the good resolving power of \mgb\ only for $\Gamma$
and large insensitivity to all remaining parameters. Different
values of [$X_{el}$/Fe] ([Ti/Fe] in this case) have in practice no
effect. Furthermore, the age and metallicity are nearly
degenerate. The open circles are the simulations, the filled
squares are the data by \citet{Longhetti20}, and finally the open
triangles are those by \citet{Trager97}. Secondly, we re-examine
in the same plane the simulations shown in Fig.~\ref{enh_simul},
in which a wider age rage for the old component is adopted. All
remaining ingredients are the same as above. These model are shown
in the right panel of Fig.~\ref{Mgb_MgFe_simul}. Comparing data
with theory, we would conclude that in both cases $\Gamma$ in the
range 0 to 0.35 and $Z\leq$0.07, likely 0.05, are best suited, the
only difference between the two type of models is that in presence
of a burst the left-lower corner of the panels is populated (i.e.
\mgb$\leq$2 and \MgFe$\leq$3). In any case the large dispersion of
the bulk population in the \Hbeta\ vs. \MgFe\ plane is once more
compatible with all galaxies being old and spanning a large range
of $\Gamma$s and $Z$s. The large age dispersion (say from 13 to
5\,Gyr) and the occurrence of a later burst of activity are no
longer unavoidable results and/or hypotheses. Only a few galaxies
seem to require a later burst of star formation.

However, the upper limits suggested by the \mgb\ vs. \MgFe\ plane are
not firmly established as different conclusions would be derived by
looking at another diagnostic plane such as \MFe\ vs. \mgii\ shown in
Fig.~\ref{Fe_Mg2_simul}. All the symbols have the same meaning as in
Fig.~\ref{Mgb_MgFe_simul}. From this diagram one would indeed conclude
that $\Gamma$ falls in the range 0.35 to 0.5, that high metallicities
(up to $Z$=0.07) are likely to occur, and finally that galaxies with
$\Gamma$=0 and metallicities larger than solar or so are not
there. Since \mgii\ seems not to depend on $\Gamma$ and \MFe\ is
likely more sensitive to both $\Gamma$ and $Z$ than other indices, we
would favor the hints arising from this diagnostic plane with respect
to the previous ones.

\noindent
{\bf Two Template Galaxies}.
It may be worth of interest here to closely inspect some of the
galaxies in the samples paying major attention to those with unusually
high \Hbeta\ and/or very low \MgFe. To this aim three sources have
been used: \citet{Trager97,Longhetti20} for the indices and central
velocity dispersions, \citet{Burstein97} for colors, masses,
luminosities. The data are listed in Table~\ref{selected}. We have not
considered RR24a and RR24b because of their negative \Hbeta.

\begin{figure}
\centerline{
\psfig{file=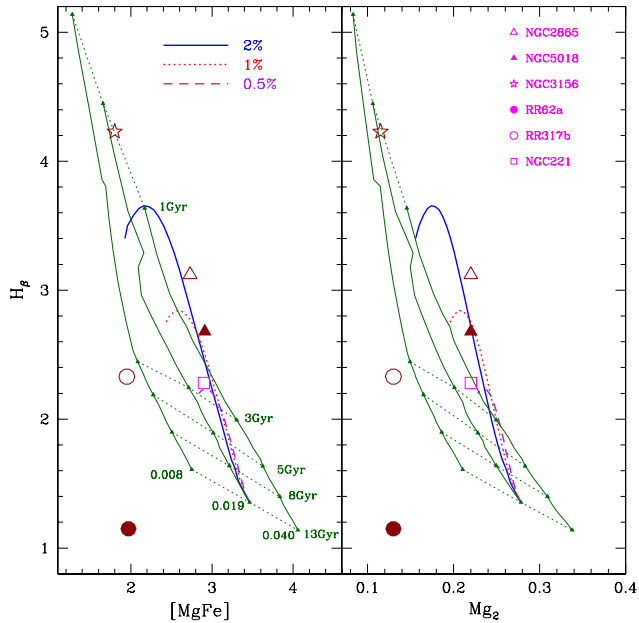,width=9.0truecm}}
   \caption{Age estimate of the last episode of star formation in a
   few galaxies with unusual values of \Hbeta\ and/or \MgFe\ (Left
   panel) and \mgii (Right Panel). The galaxies indicated with
   different symbols are listed in the Right Panel. In each panel we
   plot three SSPs with different metallicity, $Z$=0.008, 0.019, and
   0.040 as indicated, along which five values of the age are marked,
   i.e. 13, 8, 5, 3 and 1\,Gyr, and the corresponding lines of
   constant age are drawn (thin dotted lines). Finally, we show the
   locus expected for composite galaxies, in which to a population
   with the age of 13\,Gyr and metallicity $Z$=0.019, an burst of
   arbitrary age and different intensity is added. The metallicity of
   the young component is however the same as the old one
   ($Z$=0.019). The intensities are $\beta_{2}$=0.005 (long dashed
   line), 0.01 (heavy dotted line), and 0.02 (solid line). Along each
   curve the age increases from the minimum plotted value 0.1\,Gyr to
   13\,Gyr. The position of M32 is compatible with various
   combinations of $T_{B}$ and $\beta$, for instance
   $T_{B}\simeq$1\,Gyr and $\beta_{2}$=0.005 or $T_{B}\simeq$2-3\,Gyr
   and $\beta_{2}$ in the range 0.01 to 0.02. At given $\beta_{2}$,
   the burst age will increase if higher metallicities for the young
   component are considered.}
\label{5018_m32}
\end{figure}

Without going into a detailed discussion of all these objects, we may
still make several points:

(i) Adopting (B--V)=1.00$\pm$0.05 as the typical mean value for
early-type galaxies \citep{Burstein97}, those with strong \Hbeta\
tend to be bluer than normal galaxies by $\Delta$(B--V)=0.05 to
0.10 up to the extreme case of NGC~3156 with 0.28. Furthermore
there seems to be a correlation between (B--V) and \Hbeta: (B--V)
gets redder at decreasing \Hbeta.

(ii) Looking at the time dependence of (B--V) (other blue colors have
the same behavior) shown in Fig.~\ref{simul_plane} owing to the very
fast decline past the occurrence of the burst, colors in the range 0.8
to 0.9 are recovered after 1.5\,Gyr or so (the exact value depends on
the intensity of the burst), which is the age of the young component

(iii) From the position of a galaxy suspected to be in the burst
or post burst activity because of its unusual indices in planes
for composite systems like those presented in Fig.~\ref{comp_ind}
on one hand we may get another estimate of the burst age, on the
other hand we may check the consistency of the whole picture. For
the sake of illustration, we discuss here the case of NGC~5018, a
giant elliptical galaxy with total mass M=1.24$\times
10^{11}$\Msun, luminosity $L_{B}=5.43\times 10^{10}$\Lsun, central
velocity dispersion $\Sigma_{0}$=223 km/s, \Hbeta=2.30, and
\mgii=0.211. Long ago, \citet{Bertola93} have classified NGC~5018
as a giant, chemically unevolved galaxy consisting of M32-like
stars, and \citet{Leonardi00} have suggested that the central
regions of NGC~5018 are dominated by an intermediate-age
population. Along the same vein, is the very recent study by
\citet[][ private communication]{Buson03} who have stressed the
close similarity between M32 (NGC~221) and NGC~5018. The
similarity concerns the indices as shown by the entries of
Table~\ref{selected}, the (B--V) colors, and the spectrum from
2000\AA\ to 4700\AA\ \citep{Buson03}. They have also made the
point that since M32 (at least in the central regions) is commonly
considered to be dominated by a population whose age is about
3\,Gyr
\citep{Freedman89,Freedman92,Trager20a,Trager20b,Longhetti20} the
same should occur for NGC~5018. However, the reality may be more
articulated than this simple scheme. Studies of the stellar
content and other properties of M32 have pointed out that the bulk
population could be as old as that of globular clusters, e.g. see
the discussion by \citet{Bressan94} and references therein, and by
\citet{Grillmair96}. \citet{Bressan94,Bressan96} have argued that
the young stellar component in M32 could result from a recent
burst of activity superposed to the pre-existing, old stellar
content. Is NGC~5018 experiencing the same? To check this
possibility, in Fig.~\ref{5018_m32} we plot M32, NGC~5018 and
other galaxies of Table~\ref{selected} on to the \Hbeta\ vs.
\MgFe\ and \mgii\ planes and compare them with single SSPs of
different age and metallicities and composite SSPs in which bursts
of different intensity and age are superposed to an old dominant
component. In the first case we recover the standard solution,
i.e. M32 is an object whose stellar content is about 3-4\,Gyr,
whereas in the second case M32 is an old object (say 10-13\,Gyr)
on the top of which a young component (1-2\% of the total mass) is
present. The age of this is about 1\,Gyr. An old galaxy, which
underwent a burst of activity say 1\,Gyr ago, would appear as a
brand newly formed galaxy with the age of about 3-4\,Gyr. The same
analysis can be extended to other galaxies of
Table~\ref{selected}. A special remark is due to the three
galaxies with very low \MgFe: we exclude RR317b because it shows
evidence of emission [OII](3712\AA) as already pointed out by
\citet{Longhetti20}. The case of NGC~3156 is trivial because of
its strong \Hbeta: it is indeed fully compatible with theory if
undergoing a burst of star formation. We are left with RR62a for
which not very much can be said: it follows the same
$\mu_{e}$-$R_{e}$ relationship of ordinary galaxies, but it
deviates from the Hamabe-Kormendy relationship \citep{Hamabe87}.
It could be a old, low metallicity galaxy which according to
\citet{Gorgas93} should fall into the bottom left corner of the
\Hbeta\ vs. \MgFe\ plane. Alternatively it could be old galaxy
which is just starting (or undergoing) a burst of star formation
as indicated by the results for composite SSPs shown in
Fig.~\ref{comp_ind}.

\begin{table*}
\caption[]{Selected galaxies: seven indices on the Lick system;
M is the mass in solar units inside the effective radius; $L_{B}$ is
the B-luminosity in solar units; $\Sigma_{0}$ is the central velocity
dispersion in km/s.}
\begin{scriptsize}
\begin{center}
\begin{tabular*}{127mm}{|l c c c c c c c c c c r|}
\hline
\multicolumn{1}{|l}{Name} &
\multicolumn{1}{c}{\Hbeta} &
\multicolumn{1}{c}{\mgii} &
\multicolumn{1}{c}{\mgb} &
\multicolumn{1}{c}{Fe52} &
\multicolumn{1}{c}{Fe53} &
\multicolumn{1}{c}{\MFe} &
\multicolumn{1}{c}{\MgFe} &
\multicolumn{1}{c}{(B--V)} &
\multicolumn{1}{c}{$\log$M} &
\multicolumn{1}{c}{$L_{B}$} &
\multicolumn{1}{c|}{$\Sigma_{0}$} \\
\hline
\multicolumn{12}{|c|}{Galaxies with \Hbeta$\geq$2.5$\pm$0.2}\\
\hline
NGC2865      & 3.12  & 0.220 & 3.28 & 2.34 & 2.22 & 2.28 & 2.73 & 0.83 & 10.517 & 10.241 &   208\\
NGC5018      & 2.68  & 0.220 & 3.31 & 2.89 & 2.24 & 2.56 & 2.91 & 0.85 & 11.094 & 10.736 &   247\\
E2400100b    & 2.79  & 0.210 & 3.79 & 3.12 & 2.33 & 2.71 & 3.21 & 0.82 &        &        &   223\\
RR101b       & 2.35  & 0.230 & 3.67 & 2.83 & 2.35 & 2.64 & 3.08 & 0.88 &        &        &   210\\
RR187b       & 2.80  & 0.220 & 3.40 & 2.69 & 2.34 & 2.51 & 2.92 & 0.93 &        &        &   144\\
NGC2863      & 3.02  & 0.208 & 3.12 & 2.65 & 2.20 & 2.43 & 2.88 &      &        &        &    87\\
NGC4742      & 3.32  & 0.185 & 2.83 & 2.50 & 1.97 & 2.23 & 2.51 &      &        &        &   105\\
NGC5061      & 2.65  & 0.275 & 3.93 & 2.97 & 2.83 & 2.90 & 3.38 & 0.86 & 10.925 & 10.586 &   191\\
\hline
\multicolumn{12}{|c|}{Galaxies with 2$\pm$0.2$\leq$\Hbeta$\leq$2.5}\\
\hline
RR297a       & 1.97  & 0.270 & 4.08 & 2.89 & 2.32 & 2.60 & 3.26 &      &        &        &   177\\
RR387a       & 2.20  & 0.290 & 4.48 & 3.08 & 2.74 & 2.91 & 3.61 &      &        &        &   187\\
RR405b       & 1.89  & 0.340 & 4.81 & 2.72 & 2.76 & 2.74 & 3.63 &      &        &        &   207\\
RR409a       & 1.89  & 0.290 & 4.51 & 3.02 & 3.28 & 3.15 & 3.77 &      &        &        &   187\\
NGC221       & 2.18  & 0.197 & 2.93 & 2.94 & 2.53 & 2.73 & 2.83 & 0.84 &  8.592 &  8.074 &    77\\
NGC636       & 1.94  & 0.286 & 4.08 & 3.30 & 2.80 & 3.05 & 3.53 &      &        &        &   162\\
NGC1283      & 1.93  & 0.288 & 4.66 & 3.07 & 2.58 & 2.82 & 3.63 & 0.92 & 11.327 & 10.454 &   236\\
NGC1374      & 1.81  & 0.325 & 5.37 & 3.06 & 2.59 & 2.83 & 3.90 & 0.94 & 10.782 &  9.929 &   187\\
NGC1521      & 2.00  & 0.284 & 4.32 & 3.79 & 3.11 & 3.45 & 3.86 & 0.95 & 11.378 & 10.677 &   223\\
NGC1573      & 1.85  & 0.344 & 5.02 & 2.63 & 2.52 & 2.58 & 3.60 &      &        &        &   272\\
NGC1700      & 2.00  & 0.278 & 3.75 & 3.48 & 3.19 & 3.33 & 3.53 & 0.92 & 11.141 & 10.795 &   226\\
NGC2636      & 1.90  & 0.225 & 3.57 & 2.87 & 2.82 & 2.84 & 3.19 &      &        &        &    85\\
NGC2888      & 2.49  & 0.237 & 3.41 & 2.65 & 2.20 & 2.43 & 2.88 &      &        &        &    87\\
NGC3605      & 2.03  & 0.241 & 3.67 & 3.14 & 2.89 & 3.02 & 3.33 & 0.87 & 10.062 &  9.205 &   103\\
NGC3610      & 1.92  & 0.270 & 3.83 & 2.91 & 2.65 & 2.78 & 3.26 &      &        &        &   159\\
NGC3640      & 1.87  & 0.272 & 3.49 & 2.92 & 3.08 & 3.00 & 3.24 & 0.93 & 10.801 & 10.249 &   176\\
NGC3873      & 1.75  & 0.292 & 4.66 & 2.99 & 3.09 & 3.04 & 3.76 &      &        &        &   243\\
NGC4033      & 1.86  & 0.266 & 4.38 & 3.13 & 2.73 & 2.93 & 3.58 & 0.89 & 10.227 &  9.794 &   126\\
NGC4308      & 2.00  & 0.207 & 3.01 & 2.29 & 2.50 & 2.39 & 2.69 &      &        &        &    88\\
NGC4458      & 1.91  & 0.247 & 4.03 & 2.18 & 2.18 & 2.18 & 2.96 & 0.92 & 10.170 &  9.429 &   106\\
NGC4489      & 2.16  & 0.221 & 2.96 & 2.97 & 2.94 & 2.95 & 2.96 &      &        &        &    49\\
NGC4841B     & 2.01  & 0.288 & 4.80 & 3.16 & 3.54 & 3.35 & 4.01 & 0.94 & 11.485 & 10.755 &   229\\
NGC4926      & 2.10  & 0.299 & 4.82 & 3.22 & 1.95 & 2.59 & 3.53 & 0.96 & 11.417 & 10.578 &   270\\
NGC6086      & 1.93  & 0.310 & 4.87 & 3.30 & 2.71 & 3.01 & 3.83 & 0.93 & 11.936 & 10.965 &   304\\
NGC7454      & 2.09  & 0.204 & 3.24 & 2.56 & 2.12 & 2.34 & 2.75 & 0.88 & 10.512 & 10.184 &   103\\
\hline
\multicolumn{12}{|c|}{Galaxies with unusually low \MgFe\ }\\
\hline
RR62         & 1.15  & 0.130 & 2.30 & 1.57 & 1.82 & 1.69 & 1.97 &      &        &        &    81\\
RR317b       & 2.33  & 0.130 & 1.82 & 2.47 & 1.74 & 2.10 & 1.95 &      &        &        &    91\\
NGC3156      & 4.23  & 0.115 & 1.81 & 1.94 & 1.66 & 1.80 & 1.80 & 0.72 & 10.104 &  9.792 &   112\\
\hline
\end{tabular*}
\end{center}
\end{scriptsize}
\label{selected}
\end{table*}

\noindent {\bf The Coma EGs.} We have already mentioned that EGs
in Coma \citep{Jorgensen99} have systematically higher values of
\Hbeta\ than the local galaxies. The minimum values is \Hbeta=1.5.
Repeating the same analysis we have made for the \citet{Trager97}
and \citet{Longhetti20} samples we would come up with the
conclusions: (i) The $\alpha$-Model may hold good even in this
case. (ii) The Coma galaxies have a higher degree of enhancement
than the local ones. A rough estimate is $\Gamma \geq$0.3.
Supporting this idea is the estimate for the age of Coma's EGs by
\citet{Poggianti01} who find a mean age older than about 9\,Gyr;
the estimate of the ratio [Mg/Fe] by
\citet{Jorgensen97,Jorgensen99} who finds [Mg/Fe]=0.3 to 0.4 as
the central velocity dispersion $\log\Sigma_{0}$ increases of
0.4\,dex; and finally the recent study by \citet{Mehlert03} who
find \asfe\ ratios in the range 0.15 to 0.4. The mean degree of
enhancement in EGs of the Coma Cluster seems to be at least
0.1\,dex higher than in the local field galaxies.

\noindent {\bf $\alpha$-Enhancement in EGs: Data and Theory.} The
occurrence of the $\alpha$-enhancement in early-type galaxies is
currently attributed to the duration of the star forming period
and the different contribution to chemical enrichment by Type Ia
and Type II supernov\ae\ \citep[see][ for a review of the
subject]{Matteucci97}. In brief, $\alpha$-elements are mainly
ejected by Type II together with some Fe, whereas Fe is
essentially expelled by Type Ia supernov\ae\ (via the carbon
ignition in binary White Dwarfs reaching the Chandrasekhar limit
by mass accretion). As long as Type Ia supernov\ae\ do not
intervene in a significant manner, the chemical composition of the
gas and newly formed stars in turn will be enhanced in
$\alpha$-elements. Since Type Ia supernov\ae\ are expected to
start contaminating the intergalactic medium than Type II
\citep{Greggio83}, in the framework of the standard supernova
driven galactic wind model by \citet{Larson74} and the standard
initial mass function, the time scale of star formation and
galactic wind must be shorter than about 0.5\,Gyr not to decrease
the initial \asfe$>$0 (when $\alpha$-elements are mostly produced)
to \asfe$\leq$0 (when iron is predominantly ejected). In other
words, to reproduce the observed trend of the \asfe-mass
relationship, the total duration of the star forming activity
ought to scale with the galaxy mass according to $\rm \Delta
t_{SF} \propto M_{G}^{-1}$. This is the main drawback of the
standard model because it requires a mass-star-formation-timescale
correlation which is the opposite of what implied by the
Color-Magnitude relationship \citep{Bower92a} as amply discussed
in \citet{Bressan96,Chiosi00,Chiocar02} and \citet{Tantalo02}.
This point of difficulty has been overcome by the N-Body Tree-SPH
model of \citet{Chiocar02}. In brief (i) Independently of the
total mass, galaxies of high initial density undergo a prominent
initial episode of star formation followed by quiescence. (ii) The
same applies to high mass galaxies of low initial density whereas
the low mass ones undergo a series of burst-like episodes that may
stretch over a considerable fraction of their lifetime. (iii) The
mean and maximum metallicity increase with the galaxy mass.
Therefore these models can account for the CMR of EGs. (iv)
Finally, the occurrence of galactic winds does not follow the
simple \citet{Larson74} model, but takes place continuously in
lumps of gas as soon as their thermal kinetic energy exceeds the
gravitational energy of the galaxy. In other words the scheme on
which the \citet{Larson74} model rests, i.e. -- longer star
formation period -- higher metallicity -- lower degree of
$\alpha$-enhancement at increasing galaxy mass, is reversed but
for the metallicity. Therefore the CMR and $\alpha$-enhancement
constraints are met at the same time. See \citet{Chiocar02} and
\citet{Tantalo02} for all details. Given these premises, the
existence of various degrees of $\alpha$-enhancement in different
EGs finds a natural explanations thanks to the sensitivity of the
star forming process and its duration to the initial conditions
(mean density) and total mass of the proto-galaxy. By the same
token we may also explain the systematic higher abundance ratios
[$\alpha$/Fe] passing from field to cluster galaxies.

\section{What is the maximum color dispersion of EG's?}\label{colors}

Our aim here is to check at what extent a past episode of star
formation would reflect on to broad-band colors, such as (B--V),
of the host galaxy as we see it today\footnote{Preliminary
studies of the same subject are by \citet{Chiosi00} and
\citet{Chiocar02}}. To this aim we adapt the toy model of galaxy
formation we have been using in so far. The analysis below applies
both to the monolithic (isolation) and hierarchical scheme.

For the sake of simplicity let adopt a model with the minimum
number of parameters: solar scaled abundances, fixed age for the
old stellar component, here indicated as $T_{F} \simeq$12\,Gyr,
fixed mean metallicity. With no further stellar activity, passive
evolution of the galaxy would imply that at the present time
(B--V)$\simeq$1.00$\pm$0.05 whereby the uncertainty reflects the
spread in metallicity and abundance ratios.

Now suppose that at a certain age $T_{B}$ the model galaxy
undergoes an additional episode of star formation of short
duration and arbitrary intensity $\beta_{2}$ (this immediately
fixes also the intensity $\beta_{1}$ of the old component, their
sum being equal to unity). The burst can be of internal origin
(e.g. gas left over by the previous activity and then turned into
stars) or the result of a merger with some gas-rich object. In the
first case, the galaxy will be made by an old and a young
component. In the second case, the galaxy will consist of two
sub-units harboring old populations (in principle of different age
and metal content) plus the younger component born during the
burst.

\begin{figure}
\centerline{
\psfig{file=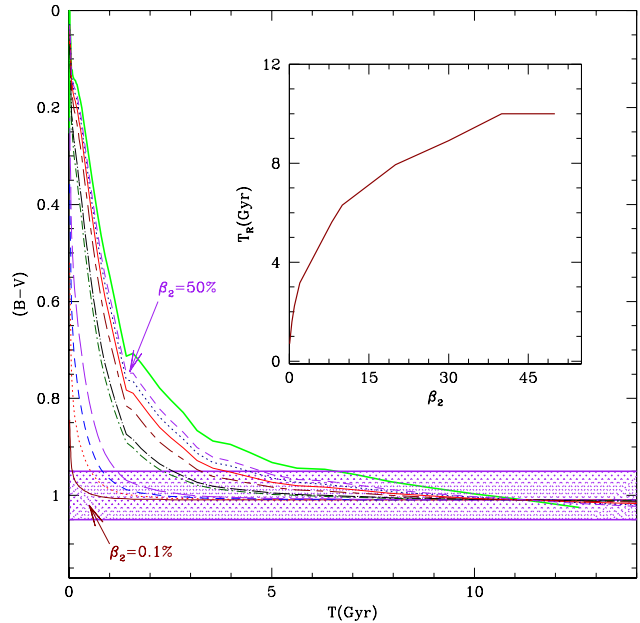,width=9.0truecm}}
   \caption{Simulations of the (B--V) color of EGs in which a burst of
   star formation of arbitrary intensity ad age $T_{B}$ is superposed
   to an old population with the age of 13\,Gyr. In these simulations
   for both components we assume $Z$=0.019 and $\Gamma$=0. The burst
   intensity is $\beta_{2}$=0.001, 0.002, 0.005, 0.01, 0.02, 0.08,
   0.10, 0.20, 0.30, 0.40, and 0.50, going from bottom-left to
   top-right as indicated. The horizontal bar visualizes the typical
   color of EGs with the uncertainty $\Delta (B-V)=\pm 0.055$. The
   insert shows the time $T_{R}$ in Gyr elapsed from the epoch of the
   burst to the stage at which the composite galaxy recovers the
   typical red colors of EGs as a function of the burst intensity.
   $T_{R}$ is given by the intersection of each color curve with the
   horizontal bar.}
   \label{mergers}
\end{figure}

The color evolution of the composite galaxy is followed up to the
present time and its broad-band colors are tested against
observations. The results are displayed in the large panel of
Fig.~\ref{mergers} as a function of the age and intensity of the
secondary activity. Let us now read off the age $T_{R}$, at which
a galaxy which underwent a burst of star formation of intensity
$\beta_{2}$ at the age $T_{B}$, is able to recover the same color
it had before the burst. This is simply given by the intersection
of any line labelled $\beta_{2}$ with the horizontal bar. The ages
$T_{R}$ as function of the $\beta_{2}$ are shown in the small
panel of Fig.~\ref{mergers}. It turns out that for many
combinations of $T_{B}$ and $\beta_{2}$, the resulting (B--V)
color would be too blue as compared to the typical colors of EG's,
(B--V)=1.00$\pm$0.05. Stellar activities engaging 5-10\% of the
total mass and taking place as early as 5-6\,Gyr ago would be
detectable. The situation gets even worse for higher $\beta_{2}$
and/or lower $T_{B}$. {\it This implies that only remote or minute
star forming events are allowed. Therefore, either mergers occur
very early on in a galaxy's history or only captures of small
bodies and little companion star formation can take place at later
epochs}. This is a rather strong constraint that should be taken
into account by any model of hierarchical assembling of big
galaxies. It is plausible to suggest that the main body of a
galaxy was assembled early on and that all subsequent activity we
infer from the observational data is limited to the capture of
small objects (likely satellites of the dominant galaxy).

\section{Summary and concluding remarks}\label{concl}

The present-day challenge with EGs is to unravel their formation
and evolution history. In a simplified picture of the issue, the
problem can be cast as follows. Do EGs form by hierarchical
merging of pre-existing sub-structures (maybe disc galaxies) made
of stars and gas? Was each merging event accompanied by strong
star formation? Or conversely, they originate from the early
aggregation of lumps of gas turned into stars in the remote past
via a burst-like episode ever since followed by quiescence so as
to mimic a sort of monolithic process? Even if the two
alternatives seem do oppose each other, actually they may concur
to shaping the final properties of EGs as seen today. The question
is to unravel the signature of the forming mechanism from the
observational data. To this aim we have examined the line
absorption indices on the Lick system of normal, field EGs of
\citet{Trager97} and of interacting EGs (pair- and shell-objects)
of \citet{Longhetti20}. Occasionally we have also looked at the
EGs in the Coma Clusters of \citet{Jorgensen99} and other more
specific cases taken from literature. The bottom line of this
study was (i) to check whether normal (quiescent) and interacting
have a different behavior in the popular diagnostic planes such as
\Hbeta\ vs. \MgFe\ (and others); (ii) to seek whether the
signature of interaction may mirror in some specific changes of
the indices that could un-equivocally hint for bursts of stellar
activity; (iii) to evaluate the intensity of those bursts or
secondary episodes of star formation; (iv) to explore whether
other alternatives can exist, i.e. distinct from obvious ones
resting on large age range and/or bursts of star formation (from
one to several) at various epochs.

From the various observational issues we have been examining in so
far, we gather the following picture:

(1) Both normal, field and interacting galaxies have the same
scattered distribution in the \Hbeta\ vs. \MgFe\ diagnostic plane even
the interacting ones show a more pronounced tail toward high \Hbeta\
values. This may suggest that a common physical cause is at origin of
their distribution.

(2) The distribution of normal and interacting galaxies is
smoothly elongated in \Hbeta. While for the interacting objects an
easy, however not fully satisfactory, explanation can be found
invoking a late burst of activity, in the case of normal galaxies
an explanation can be found only invoking a large age range for
the bulk population together with the existence of an universal
law of metal enrichment. Adding a late burst of star formation to
an old system without invoking the large age range and the
universal law of chemical enrichment would not yield the desired
trend. Therefore not only normal galaxies have to be build up over
a large time interval but also at increasing total metallicity.
The occurrence of late star forming episode is a sort of optional.

(3) More specifically, a typical model based on classical SSPs
with solar partition of elements and where the burst is
superimposed to an old and coeval population is not able to
reproduce the smooth distribution of galaxies in the \Hbeta\ vs.
\MgFe\ plane. This kind of model would predict an outstanding
clump at low \Hbeta\ values, contrary to what is observed. Models
in which star formation lasted for a significant fraction of the
Hubble time (4\,Gyr $\leq$$t_{old}$$\leq$16\,Gyr) better match the
observed diagram. In this context, the peculiar, almost {\it vertical}
distribution of galaxies (normal, shell- and pair-objects) in the
\Hbeta\ vs. \MgFe\ plane is interpreted as the trace of the
increase of the {\it average} metallicity accompanying all star
forming events. This could be the signature of a metal enrichment
happening on a cosmic scale.

(4) However, the above scheme is firstly too demanding because of
the many {\it ad hoc} ingredients that have to introduced,
secondly it neglects important effects given by the
observationally grounded hint that the stellar content of EGs is
likely enhanced in $\alpha$-elements with $\Gamma$ ranging from
0.1 to 0.4\,dex. We would like to propose a new scheme, in which
the bulk dispersion of galaxies in the \Hbeta\ vs. \MgFe\ plane is
caused by a different mean degree of enhancement. Indeed two old
galaxies of the same age (say 13\,Gyr) but different $\Gamma$ and
[$X_{el}$/Fe] for some important species (C,N,O... Ti) would have
different values of \Hbeta. For instance passing from a 13\,Gyr
old galaxy with $\Gamma$=0 to an object with the same age but
$\Gamma$=0.3 would increase \Hbeta\ by as much as 0.2-0.3 or more
(depending on the abundance ratios for some specific elements like
Ti). The effect is comparable to the mean observational
dispersion. The majority of EGs can be pretty old (10-13\,Gyr).
Furthermore, neither large age range nor universal enrichment law
are required. The {\it nearly vertical} smooth distribution for coeval
galaxies is secured by the compensatory effect of different
combinations of $Z$ and $\Gamma$. Finally, the occurrence of a
late burst of activity is not an un-avoidable ingredient of the
recipe, but a rare event interesting only those galaxies with very
high \Hbeta\ (roughly $>2.5$). The possibility that EGs span large
ranges of $\Gamma$ and metallicities but narrow ranges of ages for
the bulk population favors the monolithic scheme and can be
derived from N-Body Tree-SPH models of galaxy formation as a
consequence of the dependence of the star formation efficiency and
temporal history on the initial mean density and total mass of the
proto-galaxy.

(5) As far as we can tell, from simulations of broad-band colors,
EGs are compatible with the notion that the bulk stars have formed
in the remote past. Galaxy mergers and companion star formation in
a recent past are not likely, unless the intensity of the
secondary activity is very small (i.e. engaging less than a few
percent of the total mass). Were merging the only possible
mechanism to form massive EGs, this should have occurred in a
remote past (half of the Hubble time at least). In any case
merging of smaller units without star formation even in a recent
past cannot be excluded. Is it likely?

(6) Finally, prolonged or secondary stellar activity seem to be
also more probable in field and loose groups EGs than in those
belonging to compact groups and clusters. The physical cause could
once more be the mean density of the environment out of which
proto-galaxies are formed.

\section*{Acknowledgements}
This study has been financed by the Italian Ministry of Education,
University, and Research (MIUR), and the University of Padua under
the special contract ``Formation and evolution of elliptical
galaxies: the age problem''.

\bibliographystyle{mn2e}           
\bibliography{mnemonic,biblio}    

\end{document}